\documentclass{aa}
\usepackage{epsf}

\begin{document}

  \thesaurus{12.          
              (08.14.1;   
               02.04.1)   
            }

\title{Neutrino-pair bremsstrahlung by electrons in
                 neutron star crusts}
\author{A.D.~Kaminker\inst{1}
\and
 C.J.~Pethick\inst{2,3}
\and
A.Y.~Potekhin\inst{1}
\and
 V.~Thorsson\inst{2,4}\thanks{\emph{Present address:}
  Department of Molecular Biotechnology, University of
Washington,
Box 357350,  Seattle, Washington 98195-7350, U.S.A.
}
\and 
D.G.~Yakovlev\inst{1}
}
\institute{Ioffe Physical-Technical Institute,
              Politekhnicheskaya 26, 194021 St.-Petersburg,
        Russia
\and
 Nordita, Blegdamsvej 17, DK-2100 Copenhagen \O, Denmark
\and
 Department of Physics, University of Illinois at
        Urbana-Champaign, 1110 West Green Street, Urbana,
    Illinois 61801-3080, U.S.A.
\and
 Department of Physics, University of Washington,
Box 351560,  Seattle, Washington 98195-1560, U.S.A.
}

\offprints{C.J.~Pethick (Nordita)}

\date{Received xx September 1998 / Accepted 10 December 1998}
\titlerunning{Neutrino-pair bremsstrahlung}
\authorrunning{A.D.~Kaminker et al.}
\maketitle


\begin{abstract}
Neutrino-pair bremsstrahlung
by relativistic degenerate electrons
in a neutron-star
crust at densities $10^9$~g cm $^{-3} \la \rho \la
1.5 \times 10^{14}$~g cm$^{-3}$ is analyzed.  The processes taken
into account are neutrino emission due to
Coulomb scattering of electrons by atomic nuclei in
a Coulomb liquid, and
electron-phonon scattering (the phonon contribution) and
Bragg diffraction (the static-lattice contribution)
in a Coulomb crystal.
The static-lattice contribution is calculated
including the electron band-structure effects
for cubic Coulomb crystals of different types
and also for the liquid crystal phases composed of rod- and
plate-like nuclei
near the bottom of the neutron-star crust
($10^{14}$ g cm$^{-3} \la \rho \la 1.5 \times 10^{14}$~g cm$^{-3}$).
The phonon contribution is evaluated with proper
treatment of the multi-phonon processes which removes
a jump in the neutrino bremsstrahlung emissivity
at the melting point obtained in previous works.
Generally, bremsstrahlung
in the solid phase does not differ significantly from
that in the liquid. At
$\rho \la 10^{13}$ g cm$^{-3}$,
the results are rather
insensitive to the nuclear form factor,
but results for the solid state near the melting point
are affected significantly by the Debye--Waller factor
and multi-phonon processes.
At higher densities the Debye--Waller factor
and multi-phonon processes
become less important but the nuclear form factor
becomes more significant.
With growing $\rho$, the phonon contribution becomes
smaller.  Near the bottom of the neutron star crust
bremsstrahlung becomes less efficient
due to the reduction
of the effective electron--nucleus  matrix element
by the electron band-structure effects and the nuclear form factor.
A comparison of the various
neutrino generation mechanisms in neutron star
crusts shows that
electron bremsstrahlung is among the most important
ones.
\end{abstract}

\section{Introduction}
Neutrino-pair bremsstrahlung of electrons in
liquid and crystalline phases of dense matter is one
of the major neutrino energy-loss mechanisms in neutron star crusts.
Here, by bremsstrahlung we mean
neutrino emission due to electromagnetic
interaction of electrons with atomic nuclei.
The process can be written schematically as
\begin{equation}
 e + (Z,A) \to e + (Z,A) + \nu + \bar{\nu}.
\label{brems}
\end{equation}
It proceeds via neutral and charged electroweak currents and
leads to emission of neutrinos of all flavors.

For practical application to the thermal evolution
of neutron stars, one needs to know the neutrino
energy emission rate (emissivity) $Q$ in the density range
from about $10^{9}$ g cm$^{-3}$ to $1.5 \times 10^{14}$
g cm$^{-3}$ (the core-crust interface)
at temperatures $T \la 5 \times 10^9$ K (at which
the nuclei are not dissociated). Under these conditions,
the electrons are strongly degenerate and ultra-relativistic,
and the nuclei form either
a Coulomb liquid, or a Coulomb
crystal. For densities higher than $10^{12}$ -- $10^{13}$
g cm$^{-3}$, the melting temperature of the crystal
exceeds $5 \times 10^9$ K, and the
case of a Coulomb liquid is of no practical importance.
In the density range from about $10^{14}$ g cm$^{-3}$
to $1.5 \times 10^{14}$ g cm$^{-3}$, the nuclei
resemble rods and plates, rather than spheres
(Lorenz et al.\ \cite{lrp93};
Pethick \& Ravenhall \cite{pr95}).

The neutrino-pair bremsstrahlung process
(\ref{brems}) in a crystal is formally different
from that in a liquid. In the liquid state,
neutrinos are generated due to Coulomb
scattering of electrons by nuclei. In the solid state,
there are two contributions to the process,
electron--phonon
scattering (electron scattering by the nuclear charge
fluctuations due to lattice vibrations, referred to as the {\it
phonon contribution}),
and the Bragg diffraction of electrons, which is commonly
called the {\it static-lattice contribution}.

Neutrino-pair bremsstrahlung has been analyzed
by a number of authors
(see, e.g., Itoh et al.\ \cite{itoh-ea89}, \cite{itoh-ea96};
Pethick \& Thorsson \cite{pt97}, and references therein).
The case of a Coulomb liquid has been thoroughly
studied by Festa \&
Ruderman (\cite{fr69}), Dicus et al.\ (\cite{dcs76}),
Soyeur \& Brown (\cite{sb79}), Itoh \& Kohyama (\cite{ik83}), and,
most recently, by Haensel et al.\ (\cite{hkya96}).
The phonon contribution in the crystalline lattice
has been analyzed by Flowers (\cite{f73}),
Itoh et al.\ (\cite{itoh-ea84b}, \cite{itoh-ea89}), and also by
Yakovlev \& Kaminker (\cite{yk96}).
Notice, however, that all these authors have made
use of the so-called one-phonon approximation,
whereas, as we shall demonstrate below, multi-phonon scattering processes
can give a noticeable contribution especially near
the melting point.
The static-lattice contribution
has been considered by Flowers (\cite{f73}) and
by Itoh et al.\ (\cite{itoh-ea84a}) neglecting the electron
band-structure effects (the presence of gaps in the
electron dispersion relation).
However, Pethick \& Thorsson (\cite{pt94}, \cite{pt97}) have
shown that the band structure can
suppress strongly the static-lattice contribution and
is thus very important.
In particular, Pethick \& Thorsson (\cite{pt97})
derived a general expression for the static-lattice
contribution.

In the present article, we give an overall
analysis of neutrino-pair bremsstrahlung
by electrons in neutron star crusts, and obtain formulae
convenient for applications.
In Sect.\ 2 we describe briefly physical
conditions in neutron star crusts. In Sect.\ 3
we summarize the formalism for calculating neutrino-pair
bremsstrahlung in
the liquid and solid phases. While studying the solid phase
we consider the
ordinary body-centered-cubic (bcc) crystals, as well as
face-centered-cubic (fcc) and hexagonal-close-packed (hcp)
cubic crystals which can also
be formed in neutron-star crusts.
The phonon contribution will be calculated using a new
approach (Baiko et al.\ \cite{bkpy98}) incorporating
multi-phonon effects. It eliminates a jump in the
neutrino emissivity at the melting point obtained in
previous articles.
The static-lattice
contribution will be analyzed not only for
the traditional phase of spherical nuclei,
but also for the `exotic' phases of nonspherical nuclei.
In Sec.\ 4 we examine the main properties
of neutrino-pair bremsstrahlung
at various densities and
temperatures in neutron-star crusts, and
compare bremsstrahlung with other neutrino
emission mechanisms. We present also a simple
analytic fit for practical evaluation
of the neutrino bremsstrahlung energy loss rate.

\section{Physical conditions}
Let us outline physical conditions in a neutron
star envelope at densities
$10^9$~g~cm$^{-3} \la \rho \la  1.5 \times 10^{14}$~g~cm$^{-3}$,
and at temperatures $T \la 5 \times 10^9$ K (Sect.\ 1).
Matter in these layers consists of electrons and atomic nuclei (ions).
At densities higher than the neutron drip density,
$\rho_{\rm d} \approx 4.3 \times 10^{11}$ g cm$^{-3}$,
free neutrons appear between the nuclei
(e.g., Negele \& Vautherin \cite{nv73}).
At $\rho \ga 10^{14}$ g cm$^{-3}$,
the nuclei are likely to
form nonspherical clusters
(Lorenz et al.\ \cite{lrp93}; 
Oyamatsu \cite{o93}; Pethick \& Ravenhall \cite{pr95}).

The state of degenerate electrons is characterized by
the Fermi momentum $p_{\rm F} \equiv \hbar k_{\rm F}$
or the relativistic parameter $x$:
\begin{equation}
      p_{\rm F} = \hbar (3 \pi^2 n_{\rm e})^{1/3}, \quad
      x={p_{\rm F} \over m_{\rm e} c} \approx
      100.9 \left( {\rho_{12} Y_{\rm e}} \right)^{1/3},
\label{pF}
\end{equation}
where $Y_{\rm e}=n_{\rm e}/n_{\rm b}$ 
is the number of electrons per baryon,
$n_{\rm e}$ is the number density of electrons, 
$n_{\rm b}$ the number density
of baryons, and $\rho_{12}$
is mass density in units of $10^{12}$ g cm$^{-3}$.
In the density range under study
the electrons are ultra-relativistic $(x \gg 1)$.
The electron degeneracy temperature is
\begin{equation}
       T_{\rm F}= (\sqrt{1+x^2} -1)\,T_0, \quad
       T_0= {m_{\rm e} c^2 \over k_{\rm B}} \approx
       5.930 \times 10^9~~{\rm K},
\label{TF}
\end{equation}
where $k_{\rm B}$ is the Boltzmann constant. In our case
$T \la 5 \times 10^9$ K, and the electrons are
strongly degenerate.

\begin{figure}[ht]
\begin{center}
\leavevmode
\epsfysize=8.5cm
\epsfbox{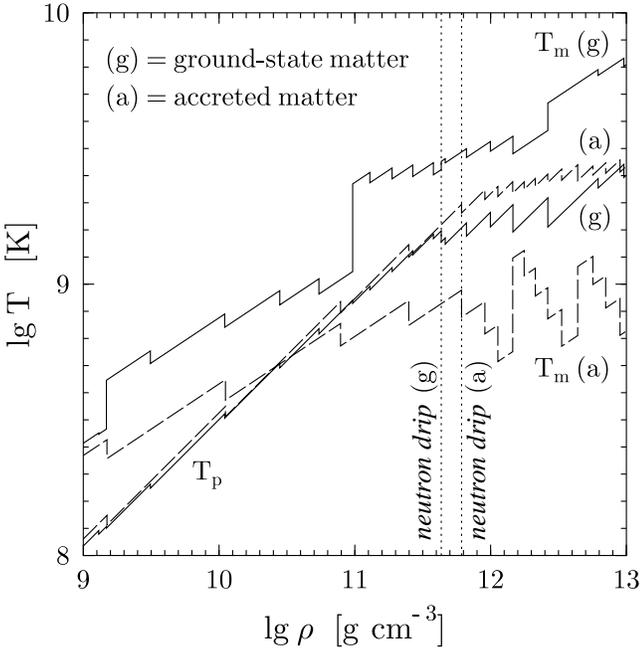}
\end{center}
\caption[ ]{
Density dependence of the melting temperature $T_{\rm m}$
and the ion plasma temperature $T_{\rm p}$, and also neutron drip
density (vertical dotted lines) for ground-state matter
and accreted matter
}
\label{figt}
\end{figure}

The nuclear composition of neutron-star envelopes
is not very well known, although it is quite certain
that light elements such as H and He
transform into heavier ones at densities
which are lower than the densities of interest.
For simplicity, we assume that only one nuclear species
is present at any fixed density (pressure).
This leads to discontinuous variations of the nuclear composition with
density (pressure). The temperature dependence of the
nuclear composition can be ignored at $T < 5 \times 10^9$ K
(e.g., Haensel et al.\ \cite{hkya96}).
For illustration, we shall make use of two models of
matter in a neutron-star crust:
ground-state (cold-catalyzed) matter and
accreted matter. For describing ground-state matter,
we shall use the following data: the results of
Haensel \& Pichon (\cite{hp94}) at
$\rho < \rho_{\rm d}$ based on new laboratory measurements
of nuclear masses with large neutron excess,
the results of Negele \& Vautherin (\cite{nv73}) for spherical nuclei at
$\rho_{\rm d} \la \rho \la 1.5 \times 10^{14}$ g cm$^{-3}$ 
derived by a
modified Hartree-Fock method, and model I of Oyamatsu
(\cite{o93}), which takes into account nonspherical nuclei.
We shall adopt the composition of accreted matter
calculated by Haensel \& Zdunik (\cite{hz90})
for $\rho \la 10^{13}$ g cm$^{-3}$, and we shall not
consider accreted matter at higher densities.
Accreted matter consists of lighter nuclei, and neutron drip
is shifted to higher density
as compared to ground-state matter (Fig.~\ref{figt}).

The state of spherical nuclei is determined by the
ion-coupling parameter
\begin{equation}
     \Gamma = {Z^2 e^2 \over a k_{\rm B} T}
     \,  \approx \, 0.225 \, x \, {Z^{5/3} \over T_8},
\label{Gamma}
\end{equation}
where $Ze$ is the nuclear charge,
$a=[3/(4 \pi n_{\rm i})]^{1/3}$
is the ion-sphere (Wigner--Seitz cell) radius,
$n_{\rm i}=n_{\rm e}/Z$ is the number density of
nuclei, and $T_8$ is the temperature in units of $10^8$ K.
For the densities and temperatures of interest,
the spherical nuclei constitute either a strongly-coupled
Coulomb liquid ($1 < \Gamma < \Gamma_{\rm m}$),
or a Coulomb crystal ($\Gamma > \Gamma_{\rm m}$), where
$\Gamma_{\rm m}$=172 corresponds
to solidification of a classical one-component Coulomb liquid
into a bcc lattice (Nagara et al.\ \cite{nnn87}).
Thus the melting temperature is
\begin{equation}
     T_{\rm m}= {Z^2 e^2 \over
     a k_{\rm B} \Gamma_{\rm m}}
     \approx 1.32 \times 10^7 Z^{5/3}
     \left( {\rho_{12} Y_{\rm e}} \right)^{1/3}~~
     {\rm K}.
\label{Tm}
\end{equation}
The density profiles of $T_{\rm m}$ for the ground-state
and accreted matter are presented
in Fig.~\ref{figt}.
The melting temperature of accreted matter
is systematically lower due to
the lower values of $Z$. If $\rho>10^{13}$ g cm$^{-3}$, one has
$T_{\rm m} \ga 5 \times 10^9$ K for
the ground-state model,
and matter is always solid for
the conditions under study.
A classical bcc lattice is bound most tightly
(see, e.g., Brush et al.\ \cite{bst66}). Therefore, it is widely assumed
that neutron-star crusts are composed of such crystals.
However, an fcc or hcp crystal is bound only slightly more weakly, so
fcc and hcp crystals may well
occur in dense stellar
matter along with the bcc ones (e.g., DeWitt et al.\ \cite{dwly93};
Baiko \& Yakovlev \cite{bya95}),
and we shall also consider this possibility.

An important parameter for a Coulomb crystal
is the ion plasma temperature
\begin{equation}
           T_{\rm p} = {\hbar \omega_{\rm p} \over k_{\rm B}}
             \approx 7.83 \times 10^9
             \left({Z Y_{\rm e} \rho_{12} \over A_{\rm i} } \right)^{1/2}~{\rm K},
\label{Tp}
\end{equation}
which is determined by the ion plasma frequency
$\omega_{\rm p}=\left( {4\pi Z^2 e^2 n_{\rm i} / m_{\rm i} }  \right)^{1/2}$,
where 
$m_{\rm i} \approx A_{\rm i} m_{\rm u}$ is the ion mass and
$m_{\rm u} = 1.66 \times 10^{-24}$~g is the atomic mass unit.
The plasma temperature characterizes
ion vibrations. If $T \ga T_{\rm p}/8$
the vibrations can be treated classically
(the high-temperature case)
while at $T \ll T_{\rm p}$ they should be treated quantum-mechanically
(the low-temperature case). In neutron-star crusts,
a transition from the high-temperature to the low-temperature
regime takes place at temperatures which
are lower than $T_{\rm m}$ (Fig.~\ref{figt}).

We shall consider ion vibrations in a purely Coulomb lattice.
For $\rho > \rho_{\rm d}$  there are neutrons in the space between ions (nuclei).
How to include the effect of these neutrons on the lattice
dynamics is at present an unsolved problem, and
we shall ignore this effect.

The effective interaction between electrons and nuclei depends on a number
of effects. One is the character of the Coulomb interaction itself,
a second is screening of the interaction by electrons, a third is the shape
of the nuclear charge distribution, and a fourth is the effect of thermal
vibrations effectively smearing out the nuclear charge, an effect taken
into account by inclusion of the Debye--Waller factor.  Thus the Fourier
transform of the effective electron-ion interaction may be
written as
\begin{equation}
        V_{\vec{q}} = {4 \pi e \rho_Z \, F(\vec{q}) \over q^2 \,
         \epsilon(q) } \, {\rm e}^{-W(q)},
\label{Vq}
\end{equation}
where $\rho_Z$ is the ion charge per unit volume,
$F(\vec{q})$ is the form factor which reflects the charge distribution
within one nucleus,
$\epsilon(q)$ is the static longitudinal dielectric
factor (Jancovici \cite{j62}), and $W(q)$ is the Debye--Waller factor.
For a crystal of spherical nuclei, we have $\rho_Z=Z e n_{\rm i}  \,$;
Eq.\ (\ref{Vq}) is valid also for non-spherical nuclei.
The Debye--Waller factor in a crystal of spherical nuclei
can be written as
\begin{equation}
   W(q) ={\hbar q^2 \over 4 m_{\rm i}} \left\langle
        {
        {\rm coth} (\hbar \omega_s / 2 k_{\rm B} T ) \over \omega_s
        } \right\rangle,
\label{DWdef}
\end{equation}
where $\omega_s$ is a phonon frequency,
$s$ enumerates phonon
modes and brackets $\langle \ldots \rangle$
denote averaging over phonon wave vectors and polarizations
(e.g., Eq.\ (8) in Yakovlev \& Kaminker \cite{yk96}).
In a Coulomb crystal $W=W(q)$ is accurately fitted by
Baiko \& Yakovlev (\cite{bya95}):
\begin{equation}
   W =
   {\alpha \over 2} \; \left( { q \over 2 k_{\rm F} } \right)^2,\quad
   \alpha = \alpha_0 \, \left( {1 \over 2} u_{-1}\,
           {\rm e}^{-9.1 t_{\rm p}} + t_{\rm p} u_{-2}
           \right),
\label{DW}
\end{equation}
where $ t_{\rm p} = T / T_{\rm p}$ and
\begin{equation}
     \alpha_0 = {4 m_{\rm e}^2 c^2 \over k_{\rm B}
           T_{\rm p} m_{\rm i}} \, x^2 \approx  1.683 \,
           \sqrt{x \over A_{\rm i} Z}.
\label{alpha0}
\end{equation}
The quantities $u_{-1}$ and $u_{-2}$ are
the frequency moments of the phonon
spectrum, $u_n = \langle (\omega_s / \omega_{\rm p})^n \rangle$.
For the bcc lattice, the frequency moments
are well known (see, e.g., Pollock \& Hansen \cite{ph73}).
For the fcc and hcp crystals, they are easily derived
by calculating the phonon spectrum from the Ewald transformation
of the lattice sums. The phonon spectra of the bcc, fcc, and hcp
lattices appear to be quite similar.
Accordingly the properties of these crystals are very similar.
In particular, $u_{-1}$ = 2.798, 2.720 and 2.703;
$u_{-2}$ = 12.972, 12.143 and 12.015, for bcc, fcc and
hcp lattices, respectively.
Note that the frequency moments for the fcc
lattice calculated by Baiko \& Yakovlev (\cite{bya95})
and used by Baiko \& Yakovlev (\cite{bya96}) and Yakovlev \& Kaminker
(\cite{yk96}) are inaccurate due to erroneous boundaries of the Brillouin
zone for this lattice implemented in the momentum averaging scheme.
Improved calculations show that Eq.\,(\ref{DW}) remains valid
with the accurate frequency moments for all the lattice types
under study. The Debye--Waller factor is important if ion vibrations
(either thermal or zero-point ones) are strong.

The simplest nuclear form factor in Eq.~(\ref{Vq})
corresponds to a spherical atomic nucleus with
a uniform proton core of radius $R_{\rm c}$:
\begin{equation}
        F(q) = {3 \over (qR_{\rm c})^3} \, \left[ \sin(qR_{\rm c})
        - qR_{\rm c} \cos(qR_{\rm c}) \right].
\label{form factor}
\end{equation}
At $\rho< \rho_{\rm d}$, the radii of the nuclei do not change under the
ambient pressure of neighbouring particles,
and one can use the standard formula $R_{\rm c} = 1.15 \, A^{1/3}$ fm.
At these densities the ratio $\eta \equiv  R_{\rm c}/a \ll 1$,
and the effect of the atomic form factor is generally negligible
($F(q) \approx 1$ for $q \la 2 k_{\rm F}$).
At densities $\rho_{\rm d} < \rho \la 10^{13}$ g cm$^{-3}$
the simplest form factor remains adequate,
but the proton core radius becomes
$R_{\rm c} \approx 1.83 \, Z^{1/3}$ fm
as deduced by Itoh \& Kohyama (\cite{ik83}) from the results
of Negele \& Vautherin (\cite{nv73}). In this case,
$\eta $ can be as high as 0.2 -- 0.3 and the
effect of the form factor is important.
For higher $\rho$, the proton charge distribution
becomes smoother, and the form factor should be
modified.

In order to describe accurately all layers of the neutron
star crusts we make use of the results
by Oyamatsu (\cite{o93}) who calculated the local
neutron (n) and proton (p) number density distributions
within a Wigner--Seitz cell and fitted them in the form
\begin{equation}
  n_j(r)= \!\left\{
     \begin{array}{ll}
       \! (n_j^{\rm in} - n_j^{\rm out})
        \left[ 1 - \left( { r \over R_j }
        \right)^{t_j} \right]^3 \!\! + n_j^{\rm out}, &  r< R_j, \\
        n_j^{\rm out},           & r \geq R_j,
     \end{array}
        \right.
\label{Oya}
\end{equation}
where $j=$n or p, and $n_j^{\rm in}$, $n_j^{\rm out}$,
$t_j$ and $R_j$ are the fit parameters.
These parameters,
as well as the sizes of Wigner--Seitz cells, are
presented  in Table 6 in Oyamatsu (\cite{o93})
for several selected values of the baryon number
density $n_{\rm b}$ for spherical and nonspherical nuclei.

In particular, Oyamatsu (\cite{o93}) gives the fit parameters
for spherical nuclei at three values of baryon number density
$n_{\rm b}$=0.01, 0.03 and 0.055 fm$^{-3}$ (i.e., $\rho=1.66 \times 10^{13}$,
$4.98 \times 10^{13}$ and $9.13 \times 10^{13}$ g cm$^{-3}$)
in the inner neutron-star crust.
These parameters are quite consistent with those
presented by Negele \& Vautherin (\cite{nv73}) for nearly the same $n_{\rm b}$.
Some of these parameters can also be deduced from Figs.\ 3 and 4
and from Table 3 of Negele \& Vautherin (\cite{nv73}) for several other
values of $n_{\rm b}$ in the inner crust. The parameters appear to
be smooth functions of $n_{\rm b}$, so
we interpolated between the
given points at $\rho_{\rm d} \leq \rho \leq 1.4 \times 10^{14}$ g cm$^{-3}$.
This interpolation smears out jumps in the nuclear composition
with increasing $\rho$, but these have little effect
on the neutrino-pair bremsstrahlung.
The interpolation allows us to calculate easily
the parameters of
spherical nuclei at any density in the inner crust. In particular,
$n_{\rm n}^{\rm out}$ gives the number density of free neutrons
outside the nuclei ($r> R_{\rm n}$) 
while $R_{\rm n}$ may be called
the nuclear radius. 
There are no protons outside nuclei, $n_{\rm p}^{\rm out}=0$,
in this regime,
and the proton core radius $R_{\rm p}$ is somewhat smaller than
$R_{\rm n}$. The nuclear mass $m_{\rm i}$ is assumed to be that of all
nucleons within $R_{\rm n}$. 
The parameters $t_{\rm n}$ and $t_{\rm p}$
range from about 4 to 6 and decrease with increasing $\rho$. Note that
the proton core radius $R_{\rm p}$ in Eq.\ (\ref{Oya}) is somewhat larger
than the proton core $R_{\rm c}$ in the simplified model (\ref{form factor}).

We have obtained also
an analytical description of atomic nuclei for lower densities,
$10^8$ g cm$^{-3} \leq \rho \leq \rho_{\rm d}$ in the outer
neutron-star crust making use of the results by
Haensel \& Pichon (\cite{hp94}).
We have adopted the same parameterization (\ref{Oya})
and constructed simple analytic expressions for the
nuclear parameters of the ground-state matter
as a function of $n_{\rm b}$ in the
outer crust. In this case $n_{\rm n}^{\rm out}=0$.
At low density in the outer crust these expressions
yield $^{56}$Fe-nuclei.

According to model I of Oyamatsu (\cite{o93}),
the phase with spherical nuclei in the inner crust
is realized up to a density 
$n_{\rm b}=0.0586$ fm$^{-3}$ ($\rho=0.973 \times
10^{14}$ g cm$^{-3}$). This is followed by the phase with
rod-like nuclei up to a density $n_{\rm b} = 0.0749$ fm$^{-3}$
($\rho = 1.24 \times 10^{14}$ g cm$^{-3}$),
and the phase with slab-like nuclei
(up to $n_{\rm b} = 0.0827$ fm$^{-3}$, $\rho = 1.37 \times 10^{14}$ g cm$^{-3}$).
Subsequently there are two phases with
the roles of nuclear matter and
neutron matter reversed, the rod-like one
(up to $n_{\rm b} = 0.0854$ fm$^{-3}$, $\rho = 1.42 \times 10^{14}$ g cm$^{-3}$),
and the ``Swiss cheese" (inverted-spheres) one, 
which is the analog of the phase with
spherical nuclei and is the last phase in the neutron-star crust
(up to $n_{\rm b} = 0.0861$ fm$^{-3}$, $\rho = 1.43 \times 10^{14}$ g cm$^{-3}$).
At higher density the nuclei dissolve to give the
uniform matter of the neutron star core.

In each crystalline phase of matter the Wigner--Seitz
cell has its own geometry, but we shall assume that in the case of bcc
phases it may be approximated by a sphere, and in rod-like phases by a
right circular cylinder.  We shall
assume that the
nucleon density distributions may be described by Eq.\
(\ref{Oya}), where
$r$ is the distance from the cell center (e.g., from
the axis of a rod in the rod-like phase,
or from the symmetry plane in the slab-like case).
We interpolate these parameters as functions of $n_{\rm b}$
within each phase separately.
In the phases with
spheres, rods and slabs, $n_{\rm p}^{\rm out}=0$,
and $n_{\rm n}^{\rm out}$ describes the number density
of free neutrons, and the region $r < R_{\rm n}$ is occupied
by the nucleus itself (with $n_{\rm n}^{\rm in} > n_{\rm n}^{\rm out}$).
In the two last ``bubble" phases with the roles of nuclear matter and
neutron matter reversed,
$n_{\rm p}^{\rm out} \neq 0$, and $n_j^{\rm out} > n_j^{\rm in}$,
i.e., the local number density of neutrons and protons
increases with distance $r$ from the center of the Wigner--Seitz
cell. With increasing density the nucleon density
profiles become smoother, approaching that for
uniform matter.

Thus we have obtained a simple analytic description
of the neutron and proton local density profiles for the
ground-state matter throughout the outer and inner
neutron-star crusts including non-spherical phases of atomic nuclei.
This description will be used below and
will be referred to as the {\it smooth
composition} (SC) model of ground-state matter.
Using this model one may
easily determine the nuclear form factor $F(\vec{q})$
numerically by calculating the Fourier
transform of $n_{\rm p}(r)$.
Unfortunately, vibrational properties
are known only for crystals composed of spherical nuclei, apart from
recent work on the elastic constants (Pethick \& Potekhin \cite{pp98}). Thus
in phases with non-spherical nuclei
we
know neither the Debye--Waller factor nor the phonon spectrum.

%
\section{General formalism}

The general expression for the neutrino 
emissivity $Q$ due to the
neutrino-pair bremsstrahlung (\ref{brems}) of relativistic
degenerate electrons in a plasma of spherical nuclei
can be written as (Haensel et al.\ \cite{hkya96})
\begin{eqnarray}
        Q &=& {8 \pi G_{\rm F}^2 Z^2 e^4 C_+^2 \over 567 \hbar^9 c^8}
           (k_{\rm B} T)^6 n_{\rm i} L
\nonumber\\ &\approx&
           3.229 \times 10^{11} \, \rho_{12} \,
           Z Y_{\rm e}  \,
           T_8^6 \, L~~{\rm erg~s^{-1}~cm^{-3}},
\label{Q}
\end{eqnarray}
where $G_{\rm F}=1.436 \times 10^{-49}$ erg cm$^3$ is the Fermi
weak coupling constant,
and $L$ is a dimensionless function
to be determined.
Furthermore, $C_+^2=C_V^2 + C_A^2+ 2(C_V^{'2} + C_A^{'2})$,
where $C_V$
and $C_A$ are the vector and axial-vector
constants of weak interaction, respectively.
We have $C_V = 2 \sin ^2 \theta_{\rm W} + 0.5$ and $C_A=0.5$,
for the emission of electron neutrinos,
to which both charged and neutral currents contribute,
and $C'_V=2 \sin^2 \theta_{\rm W} - 0.5$ and $C'_A=-0.5$, for
the emission of muonic and tauonic neutrinos, which is due only to neutral
currents. In this case, $\theta_{\rm W}$ is the Weinberg angle,
$\sin^2 \theta_{\rm W} \simeq 0.23 $. 
Equation~(\ref{Q}) with $C_+^2 \approx 1.675$
is obtained taking into account the emission of
the three neutrino flavors
($\nu_{\rm e}$, $\nu_{\mu}$, and $\nu_{\tau}$).

Thus the problem reduces to evaluating the function $L$.
In the liquid of atomic nuclei, $L=L_{\rm liq}$ is determined
by the Coulomb scattering of electrons by nuclei (Sect.\ 3.1).
In the Coulomb solid, $L$ consists of two parts,
\begin{equation}
   L= L_{\rm sol} = L_{\rm ph} + L_{\rm sl},
\label{Lsol}
\end{equation}
where $L_{\rm ph}$ describes the phonon contribution (Sect.\ 3.2) and
$L_{\rm sl}$ describes the static-lattice contribution (Sect.\ 3.3).

The non-spherical nuclei at the bottom
of the neutron-star crust form a liquid crystal at
the temperatures of interest. However, since the vibrational properties
of these phases are largely unknown, we shall not consider the phonon
contribution for this case.
The static-lattice contribution
for such a solid will be analyzed in Sect.\ 3.3.

%
\subsection{Liquid phase}
The factor $L_{\rm liq}$ is a slowly varying function
of the plasma parameters and it has basically the same
significance as the Coulomb logarithm in
calculations of transport processes in plasmas.
The most general expression for it was
obtained by Haensel et al.\ (\cite{hkya96}), who found
\begin{eqnarray}
         L_{\rm liq}& =& { \hbar c \over k_{\rm B}T}
         \int_0^{2 k_{\rm F}} \, {\rm d} q_t \,q_t^3
         \, \int_0^\infty {\rm d} q_r \,\, {S(q) |F(q)|^2 \over
         q^4 |\epsilon(q)|^2}
\nonumber\\&&\times R_T(q_t,q_r) \,
         R_{\rm NB}(q_t),
\label{Lliq}
\end{eqnarray}
where  $\vec{q}= \vec{q}_t + \vec{q}_r$ is the momentum
transfer from an electron to a nucleus in a collision event,
$\vec{q}_t$ corresponding to purely elastic Coulomb
scattering while $\vec{q}_r$
takes into account inelasticity due to the neutrino emission;
$F(q)$ is the nuclear form factor,
$\epsilon(q)$ is the static longitudinal dielectric function (Sect.\ 2),
$S(q)$ is the ion-ion
structure factor (e.g., Itoh et al.\ \cite{itoh-ea83};
Young et al.\ \cite{ycdw91}), and
$R_{\rm NB}(q_t)$ includes non-Born corrections.
The function $R_T(q_t,q_r)$ is given by Eq.\ (20) in
Haensel et al.\ (\cite{hkya96}) and describes the effects
concerned with the thermal smearing of the electron distribution
function.

In the ultra-relativistic limit, the Coulomb logarithm
depends actually on three dimensionless parameters:
    $ L_{\rm liq}=L_{\rm liq}(Z,\eta,t_{\rm F})$,
    $ \eta=R_{\rm c} /a$,
    $ t_{\rm F}= k_{\rm B} T / (2 p_{\rm F} c) \approx
    T / (2T_{\rm F})$,
where $T_{\rm F}$ is given by Eq.~(\ref{TF}).

Haensel et al.\ (\cite{hkya96}) calculated the Coulomb
logarithm (\ref{Lliq}) with the form factor
(\ref{form factor}) at $Z \leq 50$, $t_{\rm F} \la 0.1$ and
$\eta \la 0.2$
and fitted the results by an analytic formula (their Eq.\ (25)).
In this article, we shall calculate $L_{\rm liq}$
from the starting equation (\ref{Lliq}) since we shall
not restrict ourselves to the simple form factor (\ref{form factor})
(Sect.\ 2). We shall use 
the structure factor $S(q)$ obtained by 
Rogers \& DeWitt (unpublished)
and accurately fitted by Young et al.\ (\cite{ycdw91}).

Notice that if we neglect the thermal smearing of the
electron distribution, then Eq.\ (\ref{Lliq}) reduces
to the familiar expression (e.g., Festa \& Ruderman \cite{fr69})
\begin{equation}
         L_{\rm liq}=
         \int_0^{1} \, {\rm d} y
          {S(q) |F(q)|^2 \over
         y |\epsilon(q)|^2} \,\left( 1 + {2 y^2 \over 1-y^2 }
         \, \ln y  \right) 
         R_{\rm NB}(q),
\label{LliqLowT}
\end{equation}
where $y=q/(2 k_{\rm F})$. As shown by
Haensel et al.\ (\cite{hkya96}), it is a good approximation for
$T\ll\hbar c q_{\rm s}$, where $q_{\rm s} \sim a^{-1}$
is the Coulomb screening momentum.

The factor $R_{\rm NB}(q)$ represents
(Haensel et al.\ \cite{hkya96}) the ratio of the
electron scattering cross sections by atomic nucleus
calculated exactly and in the Born approximation.
It describes
the non-Born correction to the Born approximation.
To simplify consideration of the non-Born corrections
we have introduced the mean non-Born correction factor
$\bar{R}_{\rm NB}$ defined as
\begin{equation}
        \bar{R}_{\rm NB}= L_{\rm liq}^{\rm NB}/L_{\rm liq}^{\rm Born},
\label{Rnb}
\end{equation}
where $L_{\rm liq}^{\rm NB}$ and $L_{\rm liq}^{\rm Born}$
are given by Eq.\ (\ref{Lliq})
calculated with an accurate factor $R_{\rm NB}$ and
with $R_{\rm NB}=1$, respectively. We have evaluated
$\bar{R}_{\rm NB}$ using the form factor (\ref{form factor})
for wide ranges of the parameters $Z$, $\eta$, $t_{\rm F}$
(or $\Gamma$) typical for neutron star envelopes.
$\bar{R}_{\rm NB}$ appears to be a very slow function of
$\eta$ and $\Gamma$. Since our treatment of the
non-Born corrections is rather phenomenological anyway,
we have set $\Gamma=150$ and $\eta=0.1$ and
neglected the thermal smearing of the electron distribution
function. Then $\bar{R}_{\rm NB}$ is a
function of the only remaining parameter, $Z$. The numerical
results for $Z \la 60$ are accurately fitted by
%
\begin{equation}
    \bar{R}_{\rm NB}=1+0.00554 \, Z+0.0000737 \,Z^2.
\label{NonBorn}
\end{equation}
This formula enables us to calculate $L_{\rm liq}$ from
Eq.\ (\ref{Lliq}) or (\ref{LliqLowT}) with
$R_{\rm NB}(q)=1$, i.e., in the Born
approximation, and introduce the mean non-Born
correction (\ref{NonBorn}) using Eq.\ (\ref{Rnb}).

An accurate calculation of the non-Born corrections
in crystalline matter (Secs.\ 3.2 and 3.3) is a difficult
task which goes beyond the scope of the present paper.
However, we shall see (Sect.\ 4) that neutrino bremsstrahlung
in crystalline matter is quite similar to that
in a Coulomb liquid. Thus we adopt the same
factor (\ref{NonBorn}) to account
for the non-Born corrections in Coulomb crystals.

%
\subsection{Phonon contribution}
The phonon contribution in
a Coulomb crystal of spherical nuclei was studied
by a number of authors (e.g., Flowers \cite{f73};
Itoh et al.\ \cite{itoh-ea84b}; Yakovlev \& Kaminker \cite{yk96}).
So far, all articles have been restricted
to consideration of one-phonon processes
(absorption or emission of one phonon).
To allow for the
background lattice vibrations
the one-phonon reaction rate
has usually been multiplied by
${\rm e}^{-2W}$, where
$W=W(q)$ is the Debye--Waller
factor introduced in Eq.\ (\ref{DWdef}).

Under astrophysical conditions at not too
low temperatures, the main contribution
to electron--phonon scattering comes from
\emph{umklapp} processes, in which the
electron momentum transfer $\hbar\vec{q}$ in a scattering event
lies outside the first Brillouin zone. Then the
phonon (quasi)momentum is determined by reduction of
$\vec{q}$ to the first Brillouin zone.
The umklapp processes require $q \ga q_0$,
contrary to the normal processes in which
$\vec{q}$ remains in
the first Brillouin zone and $q \la q_0$,
where $q_0 \approx (6 \pi^2 n_{\rm i})^{1/3}$
is the radius of the Brillouin zone in the
approximation in which it is treated as a sphere.
Umklapp processes dominate
since the parameter
$y_0 = q_0/(2 k_{\rm F}) = (4Z)^{-1/3}$
is typically small (due to the large $Z$). Accordingly the phase
space associated with umklapp processes is much larger than
that for the normal ones (e.g., Raikh \& Yakovlev \cite{ry82})
and, in most cases, it is sufficient to consider umklapp
processes alone.

Let us rederive the expression for $L_{\rm ph}$
with a proper treatment of multi-phonon processes.
We start from the general
integral expression (Eq.~(18) of Flowers \cite{f73}) for the
neutrino emissivity due to electron-phonon scattering.
The integrand contains the inelastic part $S_{\rm d}(q,\Omega)$ 
of the dynamical 
structure factor of ions in a Coulomb 
crystal.  
It is the inelastic part that is responsible for the 
electron-phonon scattering. Baiko et al.\ (\cite{bkpy98}) 
have obtained its expression for $q \ga q_0$ by 
accurate summation of multi-phonon diagrams:  
\begin{eqnarray} S_{\rm d}(q,\Omega) & =  & 
       \int_{-\infty}^{+\infty} {\rm d}t \, {\rm e}^{i 
             \Omega t} \, S(q,t), \nonumber    \\
       S(q,t) & = & {\rm e}^{-2W} \,
        \left( {\rm e}^{\Phi(t)} -1 \right),
\label{Sdynamic}
\end{eqnarray}
where
\begin{equation}
  \Phi(t)= {\hbar q^2 \over 2 m_{\rm i}} \,
      \left\langle { \cos \left[\omega_s (t + i \hbar /2 k_{\rm B} T) \right]
           \over \omega_s  \,
          {\rm sinh} \, \left(  \hbar \omega_s / 2 k_{\rm B} T \right) }
           \right\rangle.
\label{Wd}
\end{equation}
In these calculations the density
operators are calculated to all
orders in the phonon creation and annihilation operators,
but the phonon dynamics is treated in the harmonic approximation.

We now use the approach of Flowers (\cite{f73}) with the
dynamical structure factor (\ref{Sdynamic})
and make the same simplifications as in deriving
the expression for $L_{\rm ph}$ by a semianalytical method
described by Yakovlev \& Kaminker (\cite{yk96}).
Then we obtain
\begin{equation}
         L_{\rm ph}=
         \int_{y_0}^{1} \, {\rm d} y \,
          {S_{\rm eff}(q) |F(q)|^2 \over
         y |\epsilon(q)|^2} \,\left( 1 + {2 y^2 \over 1-y^2 }
         \, \ln y   \right).
\label{Lph}
\end{equation}
In this case
\begin{eqnarray}
        S_{\rm eff}(q)&=& {63 \hbar^6 
                    \over 16 \pi^7 \, (k_{\rm B} T)^6} \,
        \int_0^\infty {\rm d}\omega \, \omega^4 \,
        \int_{-\infty}^{+\infty} \, {\rm d}\Omega \,
        \int_{-\infty}^{+\infty} \, {\rm d}t \,
\nonumber\\&&\times
        { \Omega + \omega \over {\rm e}^{\hbar (\Omega + \omega)
        /k_{\rm B} T }-1} \,
        {\rm e}^{i \Omega t} S(q,t),
\label{Sv}
\end{eqnarray}
where $\hbar \omega$ is the neutrino-pair energy.
As in Yakovlev \& Kaminker (\cite{yk96}) the lower integration
limit $y_0$ excludes the low-momentum transfers in which
the umklapp processes are forbidden.

Comparing Eqs.\ (\ref{Lph}) and (\ref{LliqLowT}) we see
that $S_{\rm eff}(q)$ plays the role of an effective static
        structure
factor that defines the phonon contribution
to the neutrino bremsstrahlung. The expression (\ref{Sv}) for it
can be easily simplified. First we can integrate over
$\Omega$ and $\omega$ which leaves us with a single
        integration over $t$:
\begin{eqnarray}
        S_{\rm eff}(q)&=& -{189 \over 2 \pi^5 } \,
        \, \left( { \hbar \over k_{\rm B} T} \right)^4 \, {\rm e}^{-2W} \,
\nonumber\\&&\times 
        {\rm Im} \, \int_{-\infty}^{+\infty} \, {\rm d}t \,
        { {\rm e}^{\Phi(t)}-1 \over
        t^5 \, {\rm sinh}^2 (\pi t k_{\rm B} T /\hbar)}.
\label{Sva}
\end{eqnarray}
The latter integration is non-trivial since
the integrand is singular at $t=0$.
However, the singularity
is easily removed by using the theory of functions
of a complex variable.
Since the function $\Phi(t)$ is analytic,
the integrand allows us
to shift the integration path into the complex $t$ plane.
The appropriate shift is $t = t' - i \hbar /(2 k_{\rm B}  T)$. It
transforms Eq.\ (\ref{Sva}) into a rapidly converging 
integral 
\begin{eqnarray}
     S_{\rm eff}(q) &=& 189 \, \left( {2 \over \pi} 
     \right)^5 {\rm e}^{-2W} \int_0^\infty {\rm d}\xi \, 
    {1 - 40 \xi^2 + 80 \xi^4 \ \over (1+4 \xi^2)^5 \, 
   \cosh^2(\pi \xi)} \, 
\nonumber\\&&\times 
     \left({\rm e}^{\Phi(\xi)} - 1 \right), \label{Sv1}
\end{eqnarray} 
where $\xi=t' k_{\rm B}  T/\hbar$ and
\begin{equation} \Phi(\xi)= {\hbar q^2 \over 2 \, m_{\rm i}}
         \left\langle { \cos (\omega_s t') \over \omega_s 
         \, {\rm sinh} \left( \hbar \omega_s / 2 k_{\rm B}  T \right)}
\right\rangle.  
\label{W_D} \end{equation}

The phonon averaging $\langle \ldots \rangle$
can be performed using the method of Mochkovitch \& Hansen 
(\cite{mh79}). Afterwards the integral (\ref{Sv1}) can be 
calculated numerically.  The behaviour of $S_{\rm eff}(q)$ 
depends on temperature, and may be characterized by the 
dimensionless parameter $t_{\rm p}=T/T_{\rm p}$ introduced in Eq.\ 
(\ref{DW}). First consider the asymptotes for $t_{\rm p} \ll 1$ 
and $t_{\rm p} \ga 1$.

In the low-temperature case, $t_{\rm p} \ll 1$, it is sufficient
to set ${\rm e}^{\Phi}-1=\Phi$ in
Eqs.\ (\ref{Sdynamic}), (\ref{Sv}) or
(\ref{Sv1}). This case corresponds
to the familiar one-phonon approximation adopted in previous
articles. In the notation introduced by Yakovlev
\& Kaminker (\cite{yk96}) one finds in this case that
\begin{equation}
      S_{\rm eff}^{\rm 1ph}(q)=
      \alpha_0 \, y^2 \, t_{\rm p} \, G(t_{\rm p}) \,{\rm e}^{-2W}
      \approx \alpha_0 \, y^2 \, b  \, t_{\rm p}^2 \,{\rm e}^{-2W},
\label{SLowT}
\end{equation}
where $\alpha_0$ is defined in Eq.\ (\ref{alpha0}), $G(t_{\rm p})$ is
the function determined by Eq.\ (12) in Yakovlev \& Kaminker (\cite{yk96}),
and $b$ is a numerical factor specified by the low-temperature
asymptote $G(t_{\rm p})=b \, t_{\rm p}$. For a bcc Coulomb crystal,
one has $b \approx 231$. This value is more accurate than
$b \approx 202$ given by the numerical fit
(15) in Yakovlev \& Kaminker (\cite{yk96}) since the latter authors
did not intend to produce a fitting formula which
would be highly accurate for $t_{\rm p} \la 0.01$.

Therefore, the results obtained in the previous articles
are strictly
valid for $T \ll T_{\rm p}$. The Debye--Waller exponent ${\rm e}^{-2W}$
included into the one-phonon approximation takes into
account renormalization of the one-phonon interaction
by background lattice vibrations.
Notice that recent calculations of $G(t_{\rm p}$) in Yakovlev
\& Kaminker (\cite{yk96}) for the fcc lattice are inaccurate due to
an error in specifying the boundaries of
the first Brillouin zone (Sect.\ 2).
After correcting this error, we obtain a result
for $G(t_{\rm p})$
very close to that for the bcc lattice.
Note also a misprint in Eq.\ (18) for the function $F_1$ in
Yakovlev \& Kaminker (\cite{yk96}): in the second term $(1.5 + \alpha)^{3/4}$
is erroneously printed instead of $(1 + \alpha)^{3/4}$
although all calculations were made using the correct expression.

In the opposite case of high temperatures,
$t_{\rm p} \ga 1$, the asymptotic form of Eq.\ (\ref{Sv1})
is very simple,
\begin{equation}
    S_{\rm eff}(q) = 1 - {\rm e}^{-2 W}.
\label{ShighT}
\end{equation}
Thus $S_{\rm eff}(q)$ becomes noticeably larger than in the one-phonon
approximation, as a result of multi-phonon processes.

We have also calculated $S_{\rm eff}(q)$ for a wide range of
temperatures $T$ and density parameters $\alpha_0$
defined in Eq.\ (\ref{alpha0}). We have verified that,
under the conditions in a neutron-star crust,
$\alpha_0 \leq 0.2$. For such $\alpha_0$ and all $t_{\rm p}$
the numerical results are fitted by a simple
expression
\begin{equation}
      S_{\rm eff}(q)= \left( {\rm e}^{2W_1}-1 \right) \,
      {\rm e}^{-2W},
\label{Sfit}
\end{equation}
where
\begin{equation}
     W_1 = { \alpha_0 \, y^2 \,  b \, u_{-2} \, t_{\rm p}^2 \over
            2\, \sqrt{ (b \, t_{\rm p})^2 + u_{-2}^2 \, \exp(-7.6 \, t_{\rm p}) }},
\label{Wfit}
\end{equation}
and $b \approx 231$, for a bcc lattice.
Notice that $W_1 \approx W \approx 0.5 \, \alpha_0 y^2 u_{-2} t_{\rm p}$
for $t_{\rm p} \ga 1$.
Notice also that $S_{\rm eff}(q)$ given by Eq.\ (\ref{Sfit})
reproduces quite accurately $S_{\rm eff}^{\rm 1ph}(q)$ 
if we replace
${\rm e}^{2 W_1}-1$ by $2 W_1$. The effect of
multi-phonon processes on neutrino bremsstrahlung
emission will be described in Sect.\ 4.

\begin{figure}[ht]
\begin{center}
\leavevmode
\epsfysize=8.5cm \epsfbox{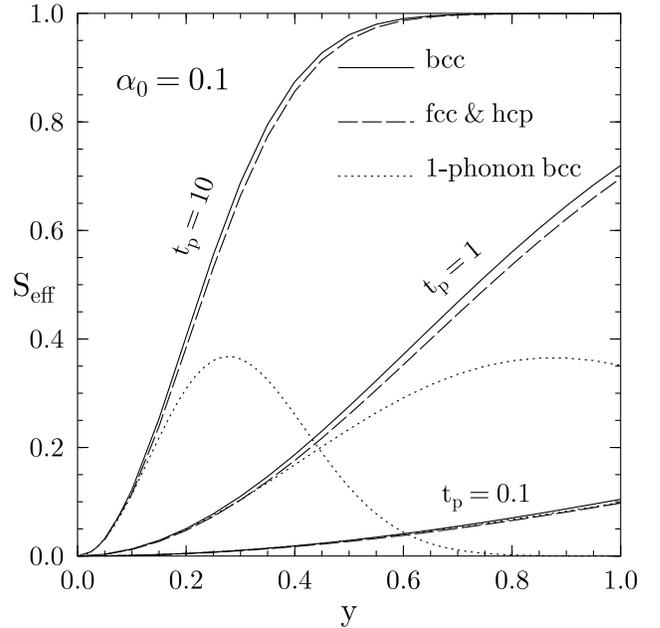}
\end{center}
\caption[ ]{
The effective structure factor $S_{\rm eff}(q)$ which enters the normalized
neutrino emissivity (\protect{\ref{Lph}}), produced by the electron-phonon
scattering, for the density parameter $\alpha_0=0.1$
and three temperatures $t_{\rm p}=$ 0.1, 1, 10 in bcc, fcc and hcp lattices.
The curves for fcc and hcp crystals almost coincide.
Dots show the effective structure factor which
corresponds to the one-phonon approximation
}
\label{figsf}
\end{figure}
Figure \ref{figsf} shows the dependence of $S_{\rm eff}(q)$
on $y$ at $\alpha_0 =0.1$ and three temperatures
$t_{\rm p}=$ 0.1, 1, and 10 for three types of Coulomb
crystals, bcc, fcc, and hcp. The results are
seen to be almost independent of lattice
type, and the curves for fcc and hcp lattices are
indistinguishable. If $t_{\rm p} \ga 1$, the one-phonon
approximation appears to be highly inaccurate for $y \ga 0.4$.

Note that the present treatment of phonon scattering
is valid as long as $T \ga T_{\rm u}$, where
$T_{\rm u} \sim T_{\rm p} \, Z^{1/3}e^2/ (\hbar v_{\rm F})$
is the temperature below which umklapp processes are frozen
out (see, e.g., Raikh \& Yakovlev \cite{ry82}), and
$v_{\rm F} \approx c$ is the electron Fermi velocity.
If $T \ga T_{\rm u}$ the electron--phonon scattering can be described
in the  free electron approximation, and
the main contribution to the scattering comes
from umklapp processes (Yakovlev \& Urpin \cite{yu80};
Raikh \& Yakovlev \cite{ry82}). The density dependence of
$T_{\rm u}$ is shown in Fig.~1 of the article by
Baiko \& Yakovlev (\cite{bya96});
$T_{\rm u}$ is so low that a study of
neutrino emission at lower temperatures is of no practical importance.

%
\subsection{Static-lattice contribution}
This contribution corresponds to neutrino
emission due to Bragg diffraction of electrons
in a crystal. It was widely assumed for a long time
that the process could be studied in the
free-electron
approximation
(e.g., Flowers \cite{f73}; Itoh et al.\ \cite{itoh-ea84a}).
However, recently Pethick \& Thorsson (\cite{pt94})
have pointed out the importance of electron
band-structure effects. The gaps in the electron
dispersion relation at the boundaries of the
Brillouin zones can reduce noticeably the static-lattice
contribution as compared to that obtained
in the free-electron approximation.
The general formalism for evaluating the static-lattice
contribution for strongly degenerate relativistic electrons
with proper treatment of band-structure
effects has been developed by Pethick \&
Thorsson (\cite{pt97}). Their Eq.\ (28) is valid for
spherical and nonspherical nuclei and can be written as
\begin{eqnarray}
        Q_{\rm sl} &=&
        {2 \pi G_{\rm F}^2 k_{\rm F} C_+^2 \over 567 \hbar^9 c^8}
        (k_{\rm B} T)^8 J \,
\nonumber\\&\approx&
        1.254 \times 10^9 \; (\rho_{12} Y_{\rm e})^{1/3}
        T_8^8 \, J~~{\rm erg~s^{-1}~cm^{-3}},
\label{Qsl}
\end{eqnarray}
where
\begin{equation}
     J =  \sum_{\vec{K}\neq0} \;
     {y^2 \over t_V^2} \; I(y,t_V)
\label{S}
\end{equation}
is a dimensionless function
given by a sum over all
reciprocal lattice vectors $\vec{K} \neq 0$
for which $\vec{K}/2$ lies within the
electron Fermi surface;
$y=K/(2 k_{\rm F})$ (with $y<1$).
The function $I(y,t_V)$
is given by Eq.\ (29) in Pethick \& Thorsson (\cite{pt97}),
which can be rewritten as
\begin{eqnarray}
       I(y,t_V) &  =  & {63 \over 8 \pi^7} \,
       {y \over y_\perp^4} \,
       \int_0^\infty \, \int_0^\infty \,
       {{\rm d}z_1 \, {\rm d}z_2 \over E_1 E_2} \,
       \int_{w_1}^{w_2} \, {\rm d}w {w^2\over
       {\rm e}^w -1}
\nonumber\\&&\times 
       [(w_2-w)(w-w_1)]^{3/2} ,
\label{Igeneral}
\end{eqnarray}
where $y_\perp = \sqrt{1-y^2}$,
\begin{eqnarray}
        E_{1,2} & =  & \sqrt{1+ \left(z_1  \mp  {z_2 \over 2} \right)^2},
\nonumber \\
        w_{1,2} & =  & \frac{E_1+E_2 \mp y_\perp
        \sqrt{(E_1+E_2)^2 -z_2^2}}{t_V y^2};
\label{Ew}
\end{eqnarray}
$t_V$ is defined in terms of the Fourier transform of the
lattice potential (\ref{Vq}) with \vec{q}=\vec{K} by
\begin{equation}
        {1 \over t_V} = \frac{| V_{\vec{K}} | y_\perp }{k_{\rm B} T} =
        \, {4 \pi e \rho_Z \over K^2 k_{\rm B} T} \;
        {| F(K) | \over | \epsilon(K)| } \,
        y_\perp \,  {\rm e}^{- W(K)},
\label{tv}
\end{equation}
and other notations were introduced in Eq.\ (\ref{Vq}).

For a lattice of spherical nuclei, we can use
Eq.\ (\ref{Q}) with
\begin{eqnarray}
     L_{\rm sl}&=&
       {\pi Z^2 \over 3 \Gamma^2} \, (9\pi Z/4)^{1/3}\, J
\nonumber\\&=&
     {1 \over 12 Z} \sum_{\vec{K}\neq0} \;
     {y_\perp^2 \over y^2} \;
     {|F(K)|^2 \over |\epsilon(K)|^2} \;
      I(y,t_V) \; {\rm e}^{-2W(K)}.
\label{Lsl}
\end{eqnarray}
According to Eqs.~(\ref{DW}), (\ref{tv}) and (\ref{Lsl}),
the Debye--Waller
factor suppresses the
electron--lattice interaction at large
reciprocal lattice vectors $\vec{K}$
and weakens the neutrino emission.
Computing  $Q_{\rm sl}$  directly
as a sum of 3D integrals (\ref{Igeneral})
is time consuming
since the number of reciprocal lattice vectors $\vec{K}$ involved is
generally large ($\sim 4Z$ terms). We simplify
computation by producing 
an analytic fit to $I(y,t_V)$.

Analytical asymptotes of $I(y,t_V)$
can be derived in the limiting
cases of high and low temperatures.
In the high-temperature limit, $t_V \gg 1$,
Pethick \& Thorsson (\cite{pt94}, \cite{pt97}) obtained
\begin{equation}
       I = {1 \over y_\perp^2 y} \left(
       1 + {2 \, y^2 \over y_\perp^2} \ln y  \right).
\label{IhighT}
\end{equation}
Inserting this asymptote into Eqs.\ (\ref{Lsl}) and (\ref{Q})
one immediately reproduces the well-known result
of Flowers (\cite{f73}) and Itoh et al.\
(\cite{itoh-ea84a}) for the static-lattice contribution when
band-structure effects are neglected.
Replacing the sum over $\vec{K}$ by an integral
over $q$, we arrive at the expression
\begin{equation}
         L_{\rm sl}^{(0)}=
         \int_{y_0}^{1} \, {\rm d} y\,
          { |F(q)|^2 {\rm e}^{-2W} \over
         y |\epsilon(q)|^2} \,\left( 1 + {2 y^2 \over 1-y^2 }
         \, \ln y   \right),
\label{Lsl1}
\end{equation}
which is similar to Eqs.\ (\ref{LliqLowT}) and (\ref{Lph})
in the liquid and for the phonon contribution in the solid,
respectively. The Debye--Waller exponent ${\rm e}^{-2W}$
is seen to play the role of the
diffraction part of an ``effective static structure factor''
that defines the static-lattice contribution (smoothed
over diffraction peaks due to replacing summation
by integration). Thus the sum
$L_{\rm ph}+L_{\rm sl}^{(0)}$ in a crystal can be written in the same
form (\ref{LliqLowT}) as $L_{\rm liq}$ in a liquid, with an effective
structure factor $S_{\rm sol}(q)={\rm e}^{-2W}+ S_{\rm eff}(q)$.
We have verified that 
$S_{\rm sol}(q)$ resembles the structure
factor $S(q)$ in a strongly coupled liquid (Young et al. \cite{ycdw91})
for ion coupling parameters $100  \la \Gamma \la 225$
if we smear out the familiar diffraction peaks in $S(q)$;
the integral contributions of the two factors are nearly the same.
This elucidates the similarity of neutrino-pair
bremsstrahlung in a liquid and a crystal (Sect.\ 4).

The asymptote (\ref{IhighT})
is temperature-independent.
The lowest-order
thermal correction can be taken into account
by introducing the factor $[1+(63/40)(\pi t_V y)^{-2}]$.

The low-temperature asymptote for $t_V \ll y_\perp /y^2$ is
(Pethick \& Thorsson \cite{pt97})
\begin{eqnarray}
       I(y,t_V)&=& {189 \over 2 \pi^{11/2} (1-y_\perp)^{1/2}
       (1+y_\perp)^2 \, (y_\perp t_V)^{5/2}}  \,
       \,
\nonumber\\&&\times 
       \exp \left(- {2  \over t_V (1+y_\perp)} \,
        \right),
\label{IlowT}
\end{eqnarray}
while for $y_\perp/y^2  \ll  t_V \ll 1$ the latter asymptote
has to be multiplied by
$\sqrt{\pi}[y_\perp/(t_V y^2)]^{5/2}$.

We calculated $I(y,t_V)$ from Eq.\,(\ref{Igeneral}) numerically
for wide ranges
of $y$ and $t_V$ and derived an analytic fit which
reproduces numerical results and the asymptotes:
\begin{eqnarray}
        I(y,t_V) & \!\! = &\!\!
        {0.3088  \, (1 +  8.416  y_{\perp} t_V)
        \over y \, (1+y_\perp)^{3/2} \,
        (y_\perp t_V)^{5/2}\, D} \,
        \exp \!\left[ - {2 \over  t_V (1+ y_\perp)} \right] \,
   \nonumber \\
        & + &
        \frac{1}{y_\perp^2 y} \,
        \left( 1 +
        \frac{0.4031}{t_V^2 y^2 +
        0.5 t_V + 0.2678} \right)
\nonumber\\&&\times 
        \left( 1 + {2y^2 \over y_\perp^2}
        \ln y \, \right) \exp \left( - {2 \over t_V H} \right),
\label{Ifit}
\end{eqnarray}
where
$D =  u^5 + 0.7124 \, u^4 -
1.689 \, u^3 + 5.237 \, u^2 - 0.2 \, u + 1.772$,
$H= 1 + y_\perp + 8.212 \, t_V y^2$,
and $u=y \sqrt{t_V/y_{\perp}}$.
The mean error of the fit is $2\%$, and the maximum
error of $7.4\%$ occurs at $t_V=0.045$ and
$y =0.987$. The fit permits a rapid and accurate
evaluation of the static-lattice contribution from Eqs.\
(\ref{Qsl}) and (\ref{S}) or  (\ref{Q}) and (\ref{Lsl})
for any plasma parameters of practical interest.

For a lattice of spherical nuclei, we can use
Eq.\ (\ref{Lsl}) with the Debye--Waller
factor and the nuclear form factor, and
calculate the sum over reciprocal lattice vectors for
bcc, fcc and hcp crystals.
In the case of non-spherical
nuclei, we use the more general Eqs.\ (\ref{Qsl}) and (\ref{S})
including the form factor but
setting $W=0$,
since the Debye--Waller factor
is unknown.
The sum over reciprocal lattice vectors in Eq.\ (\ref{S})
for a lattice of non-spherical nuclei is different from 
that for ordinary crystals.  In the case of rod-like 
phases, the lattice is the simplest two-dimensional 
triangular one, for slab-like nuclei the lattice is purely
one-dimensional, and
in the case of the Swiss-cheese phase with neutron drops immersed in
nuclear matter we shall assume that the lattice is bcc.

%
\section{Results and discussion}
Let us outline the main properties of neutrino-pair
bremsstrahlung by relativistic degenerate electrons
in neutron-star crusts. The results of Sect.\ 3 allow us
to calculate the neutrino emission for bcc, fcc and hcp
Coulomb crystals of atomic nuclei. However, in all three cases
the emission appears to be practically the same
because of the close similarity of the crystals.
For simplicity
we therefore consider bcc crystals throughout this section.
\begin{figure}[ht]
\begin{center}
\vspace*{1ex}
\leavevmode
\epsfysize=8.5cm \epsfbox{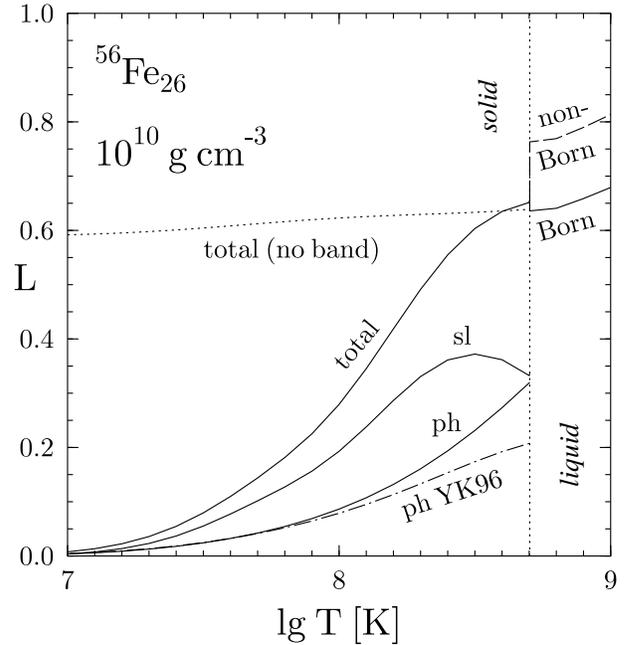}
\end{center}
\caption[ ]{
Temperature dependence of the
normalized neutrino emissivity $L$ for iron matter
at $\rho = 10^{10}$ g cm$^{-3}$. Solid lines:
Born results for the liquid phase;
phonon and static-lattice contributions, as well as
the total function (\protect{\ref{Lsol}}) for the
crystalline phase.
Dotted line: the total function for crystalline phase
but without band-structure effects.
Dashed line: non-Born result in the liquid phase.
Dot-dashed line: one-phonon approximation for the phonon
contribution (Yakovlev \& Kaminker \cite{yk96}).
All curves but one in the liquid phase are obtained
in the Born approximation
}
\label{figfe}
\end{figure}
\begin{figure}[ht]
\begin{center}
\vspace*{1ex}
\leavevmode
\epsfysize=8.5cm \epsfbox{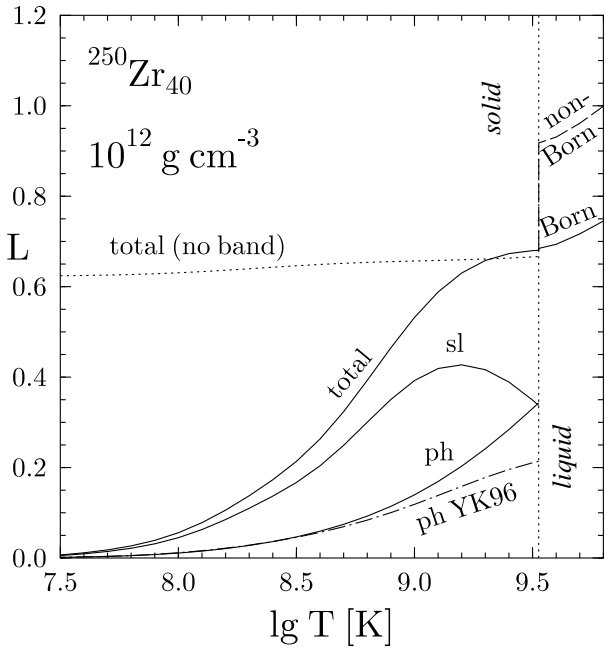}
\end{center}
\caption[ ]{
Same as in Fig.~\protect{\ref{figfe}} but for
ground-state matter composed of ${}^{250}_{\;40}$Zr nuclei
at $\rho=10^{12}$ g cm$^{-3}$
}
\label{figzr}
\end{figure}

Figures \ref{figfe} and \ref{figzr} show the temperature
dependence of
the normalized neutrino emissivity $L$ for two densities,
$\rho = 10^{10}$ and $10^{12}$ g cm$^{-3}$, in the outer and
inner crusts, respectively.
Figure \ref{figfe} is plotted for matter composed of iron
($Z=26$, $A=56$), while Fig.~\ref{figzr} corresponds to the
ground-state matter composed of
\element[][250][40]{Zr} nuclei ($Z=40$, $A=250$,
where $A$ is the number of
nucleons per Wigner--Seitz cell,
see Negele \&
Vautherin \cite{nv73}).  Vertical dotted lines separate
liquid and solid phases. The upper (dashed) line in the liquid phase
is obtained from Eq.\ (\ref{Lliq})  with the
non-Born corrections included in the function
$R_{\rm NB}$ (Haensel
et al., \cite{hkya96}). The lower (solid) line is also obtained
from Eq.\ (\ref{Lliq}) but in the
Born approximation ($R_{\rm NB}=1$).
Solid lines in the crystalline phase
show the phonon contribution (Sect.\ 3.2), the static-lattice
contribution (Sect.\ 3.3), and the total function $L_{\rm sol}$ 
given by Eq.~(\ref{Lsol}).
The dotted line also gives the total function
but neglecting the band structure effects
in the static-lattice contribution (by using the asymptote
(\ref{IhighT}) in Eq.\ (\ref{Lsl})).
Finally, the dot-dashed line
displays the phonon contribution calculated in the 
one-phonon approximation adopted in previous articles 
(Sect.\ 3.2).

The temperature profiles of
$L$ in Figs.\ \ref{figfe} and \ref{figzr} are
similar. The phonon contribution is generally
several times smaller than the static-lattice one.
Each term in the static-lattice sum (\ref{Lsl})
is suppressed exponentially with decreasing temperature
but the sum itself decreases more like a power law because, for
the smallest reciprocal lattice vectors $|\vec{K}|$,
the exponential decrease of the contribution
starts to operate
at much higher temperature than that for larger reciprocal lattice
vectors (Pethick \& Thorsson \cite{pt97}). At very low temperatures, the
reduction of contributions from all reciprocal lattice vectors
becomes exponential, and the total 
static-lattice contribution is suppressed exponentially.  
However, such low
temperatures are of no practical
importance. Generally, the static-lattice neutrino emission
is usually partially suppressed
by band-structure
effects, and these effects are substantial.

All calculations of neutrino-pair emission in
the crystalline phase have been made in the Born approximation.
On the other hand, Haensel et al.\ (\cite{hkya96}) calculated
$L_{\rm liq}$ beyond the Born approximation. For comparison of
the results in the crystalline and liquid phases,
we present also the Coulomb logarithm
$L_{\rm liq}$ determined in the Born approximation.
This curve matches the function $L_{\rm sol}$ in the solid
phase much better and makes $L$ an almost continuous
function of temperature at the melting point.
We believe that the non-Born curve in the solid phase
(which is difficult to calculate exactly) would match equally well
the non-Born curve in the liquid phase,
because of the similarity of the bremsstrahlung
in a liquid and in a crystal, mentioned in Sect.\ 3.3.
The state of a Coulomb system
(liquid or solid) is expected to have little effect on
neutrino-bremsstrahlung, since in both cases neutrino emission
occurs due to scattering of electrons by fluctuations of
electric charge produced by ions (nuclei). In the solid phase
near the melting point
both
the phonon and the static-lattice contributions must be included
for the total emission rate to match the
results in the liquid phase.
The one-phonon approximation is seen to be
generally quite accurate at low temperatures, $T \ll T_{\rm m}$
(actually at $T \ll T_{\rm p}$, as discussed in Sect.\ 3.2)
but underestimates the phonon contribution near the
melting point. It is the proper inclusion of multi-phonon processes
that makes the phonon contribution larger and almost
removes the jump of the total neutrino emissivity
at the melting point. Notice that the phonon and 
static-lattice contributions at the melting become nearly 
equal.  The physical properties of a Coulomb liquid and a 
Coulomb solid near the melting point are nearly the same. 
To verify this statement we have shifted artificially the 
melting temperature (which actually corresponds to $\Gamma 
= \Gamma_{\rm m} =172$, Sect.\ 2).  We have considered the cases 
of a supercooled liquid by taking $\Gamma_{\rm m}
= 225$, for which the liquid-state structure factors of 
Young et al.\ (\cite{ycdw91}) are available, and the case of a 
superheated crystal by taking $\Gamma_{\rm m} =100$.
In all the cases the total normalized neutrino emissivities 
$L$ do not differ noticeably from those obtained at 
$\Gamma_{\rm m} = 172$, and the discontinuity of $L$ at the 
melting point is minor.

\begin{figure}[t]
\begin{center}
\leavevmode
\epsfxsize=8.5cm \epsfbox{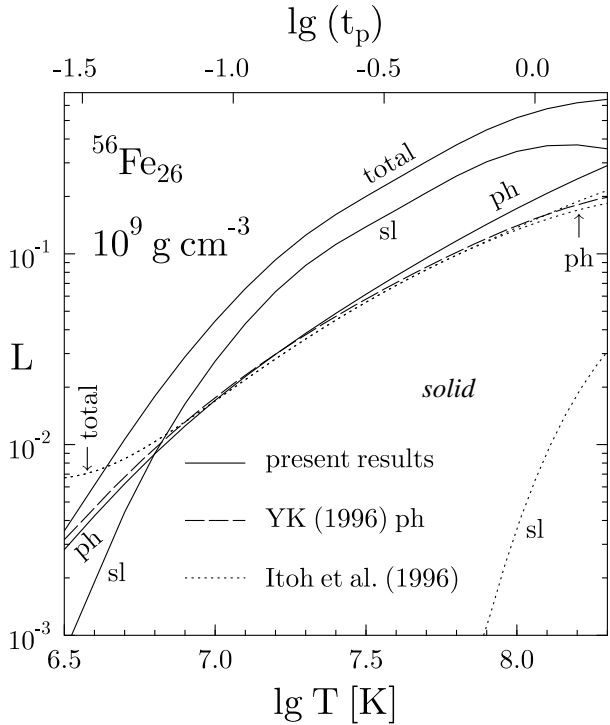}
\end{center}
\caption[ ]{ Phonon (ph), static-lattice (sl) and total
normalized neutrino emissivities
$L_{\rm ph}$, $L_{\rm sl}$, $L_{\rm sol}$
vs $T$ (lower horizontal scale) or $t_{\rm p}$ (upper horizontal scale)
for a crystal of $^{56}$Fe nuclei
at $\rho= 10^9$ g cm$^{-3}$ calculated in the Born approximation.
Solid curves show the present results,
dashes show the phonon contribution in the
one-phonon approximation (Yakovlev \& Kaminker \cite{yk96}) while dots are
the results of Itoh et al.\ (\cite{itoh-ea96})
}
\label{figfe9}
\end{figure}

\begin{figure}[ht]
\begin{center}
\leavevmode
\epsfxsize=8.5cm \epsfbox{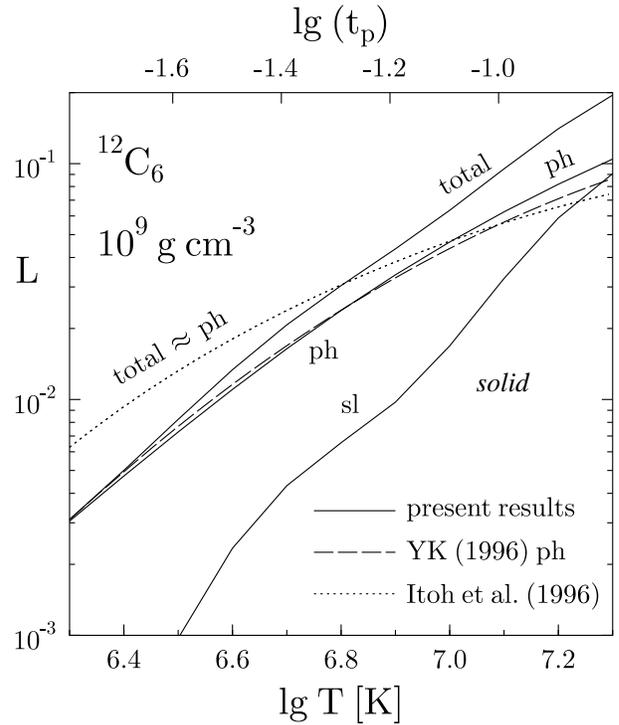}
\end{center}
\caption[ ]{
Same as in Fig.~\protect{\ref{figfe9}} but for carbon
crystal}
\label{figc9}
\end{figure}

Figure \ref{figfe9} compares the present
results for $^{56}$Fe matter at $\rho=10^9$ g cm$^{-3}$
with the familiar results by N.\ Itoh
and his group reviewed
recently by Itoh et al.\ (\cite{itoh-ea96})
and with the results by Yakovlev \& Kaminker (\cite{yk96})
for the phonon contribution obtained in the one-phonon
approximation.\footnote{%
Note that Yakovlev \& Kaminker (\cite{yk96}), while
comparing their results for the one-phonon contribution
in the iron and carbon plasmas at $\rho = 10^9$ g cm$^{-3}$
(their Figs.\ 3 and 4) with
the results of Itoh et al.\ (\cite{itoh-ea96}),
inaccurately plotted (by long dashes)  the fit
expressions of Itoh et al.\ (\cite{itoh-ea96}). The correct curves are
closer to the results by Yakovlev \& Kaminker (\cite{yk96}) and
are plotted in the present Figs.\ \ref{figfe9} and \ref{figc9}.
Two of the authors (DY and AK) are grateful
to Prof.\ N.\ Itoh for pointing out this omission.
}
The temperatures displayed are rather
low, so the one-phonon approximation almost coincides
with the multi-phonon calculation.
One can see a transition from
power-law to exponential decrease in our static-lattice curve
with decreasing temperature at $ T \sim 10^7$ K.
For lower $T$, the phonon contribution dominates over
the static-lattice one. The static-lattice contribution
given by Itoh et al.\ (\cite{itoh-ea96}) is
underestimated by several orders of magnitude.
The authors calculated this contribution
neglecting the electron band structure and
multiplied this result by a factor which
should approximately take into account the suppression
of the neutrino emission
due to band structure.
The latter factor
was chosen on the assumption
that one particular reciprocal lattice vector gave the dominant
contribution at all temperatures.
If this were the case, the suppression would be exponential.
However, as shown above, the suppression of contributions from the
various reciprocal lattice vectors sets in
at different temperatures, and the resulting suppression of the total
rate is much weaker than given by the
approximation of Itoh et al.\ (\cite{itoh-ea96}). In fact,
the approximate suppression
factor introduced by Itoh et al.\ (\cite{itoh-ea96}) makes the 
static-lattice contribution negligible at all densities and
temperatures of practical interest. Since the actual static-lattice
emission is commonly several times larger than the
phonon one, the approach of Itoh et al.\ (\cite{itoh-ea96})
systematically underestimates the total neutrino emissivity
in the solid phase of dense matter.
In addition, the phonon contribution given
by Itoh et al.\ (\cite{itoh-ea96})
appears to be overestimated
for $T \la 10^7$ K (Fig.~\ref{figfe9})
as a result of the
analytic fits proposed by Itoh et al.\ (\cite{itoh-ea96}) being insufficiently
accurate
(Yakovlev \& Kaminker \cite{yk96}).

%

Figure \ref{figc9} presents a similar comparison of the
results but for carbon.
The case of carbon at high density
is extreme since zero-point vibrations of the light
carbon ions become very strong. Nuclear reactions and
beta captures tend to transform carbon into heavier
elements. The Debye--Waller factor is
very large owing to zero-point vibrations
($\alpha \propto u_{-1}/\sqrt{A_{\rm i}Z}$ at $t_{\rm p} \ll 1$, 
see Eq.~(\ref{DW})).
It suppresses drastically the static-lattice
contribution and makes it generally smaller than
the phonon contribution. The simplified treatment
of the band-structure effects by Itoh et al.\ (\cite{itoh-ea96})
damps the static-lattice contribution especially strongly,
by several orders of magnitude,
making  $L_{\rm sol} \approx L_{\rm ph}$.

\begin{figure}[ht]
\begin{center}
\vspace*{1ex}
\leavevmode
\epsfysize=8.5cm \epsfbox{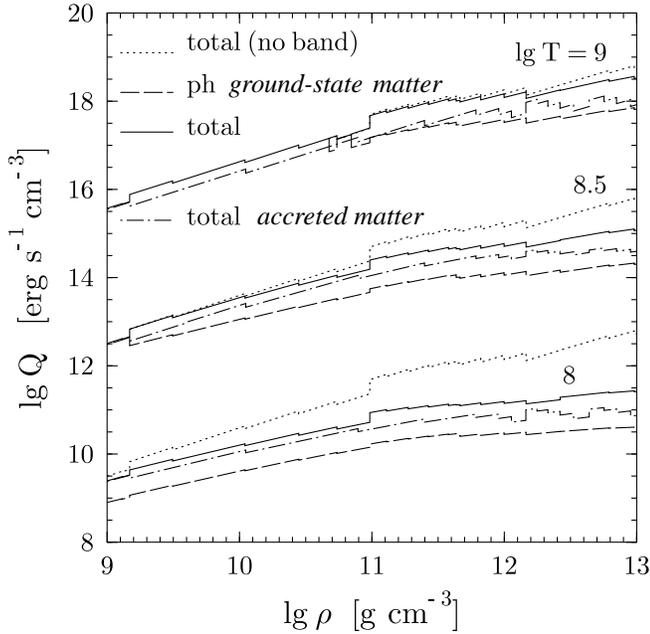}
\end{center}
\caption[ ]{
Density dependence of the neutrino bremsstrahlung emissivity
at $T=10^8$, $10^{8.5}$ and $10^9$ K for ground-state and
accreted matter calculated with the form factor
(\protect{\ref{form factor}}).
Solid and dashed lines show the total and
phonon emissivities, respectively, for ground-state matter;
dots present the same total emissivity but obtained neglecting
the electron band-structure effects. Dot-dashed lines
display the total emissivity for accreted matter
}
\label{figgro}
\end{figure}
%

Figure \ref{figgro} shows the density dependence
($10^9 $g cm$^{-3} \leq  \rho \leq  10^{13}$ g cm$^{-3}$)
of the neutrino emissivity
at three values of $T$ for the ground-state
and accreted matter. Here and in all subsequent figures
the emissivities are calculated in the Born approximation
and multiplied by the non-Born correction factor (\ref{NonBorn})
as discussed in Sect.\ 3.1.
Note that self-consistent models
of accreted matter
(e.g., Miralda-Escud\'{e} et al.\ \cite{mhp90})
correspond to $T \sim 10^8$ K. We use the accreted model
for higher $T$ to illustrate how variations of nuclear
composition affect the neutrino emission. For a particular temperature and
density, matter will be either liquid or solid.
In the liquid state, we calculate $Q$ using Eq.\ (\ref{Lliq}).
In the solid state, we present the total and phonon neutrino
emissivities for ground-state matter and the total
emissivity for the accreted matter.
In accreted
matter
the ratio of the
total emissivity to the phonon one
is qualitatively the same as in ground-state
matter. We show also the total neutrino emissivity for
ground-state matter
neglecting band-structure effects.
In the displayed density range, matter is
entirely solid for $T=10^8$ K;
there is one melting point for $T=10^{8.5}$ K
($\lg\rho_{\rm m}~[{\rm g\,cm}^{-3}]=9.17$, for ground-state matter)
which separates liquid (at $\rho < \rho_{\rm m}$)
and solid (at $\rho > \rho_{\rm m}$);
and there is a series of melting points at $T=10^9$ K
due to the non-monotonic behaviour of the melting curves
$T_{\rm m}=T_{\rm m}(\rho)$ associated with
strong variations of the nuclear composition.
The positions of the melting points can be
deduced from Fig.~\ref{figt}. For ground-state matter,
these positions can also be traced in Fig.~\ref{figgro}
from the appearance of the
phonon contribution (dashed lines).
With decreasing temperature
the solidification front shifts toward lower densities.
The accreted matter solidifies at higher densities than 
does ground-state matter due to the lower atomic number. 
At the melting points the neutrino emissivities exhibit 
some jumps, but these are small because, as we remarked 
earlier, neutrino bremsstrahlung does not change 
qualitatively while passing from liquid to solid matter.
Other, stronger jumps are associated with variations of the 
nuclear composition (Sect.\ 2). Notice that the jumps of 
both types may be ignored in practical applications.  The 
reduction of the neutrino emission by the band-structure
effects becomes stronger with decreasing temperature
and reaches one order of magnitude
for $T \sim 10^8$ K and $\rho \ga 10^{11}$ g cm$^{-3}$.
The band-structure reduction is power-law
(non-exponential) for the parameters displayed in Fig.~\ref{figgro}.
The ratio of the phonon contribution to the static-lattice one
remains nearly constant for a wide range of temperatures much below
the melting temperature, and the static-lattice contribution
is several times larger than the phonon one.
The neutrino bremsstrahlung in the accreted matter is
lower than in the ground-state
matter due to the lower atomic number, but the
        difference is not
large.

\begin{figure}[ht]
\begin{center}
\vspace*{1ex}
\leavevmode
\epsfysize=8.5cm \epsfbox{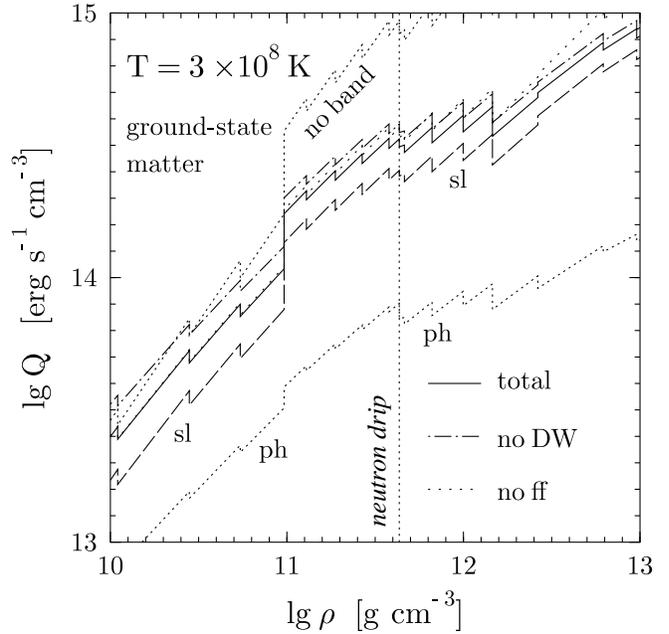}
\end{center}
\caption[ ]{
Density dependence of the neutrino bremsstrahlung
emissivity at $T=3 \times 10^8$ K for ground-state
matter calculated with the form factor
(\protect{\ref{form factor}}). Solid line:
total emissivity.
Also shown are: static-lattice contribution (sl),
phonon contribution (ph)
and the total contributions obtained neglecting either
band-structure effects (no band) or Debye--Waller factor (no DW)
or nuclear form factor (no ff)
}
\label{figcomp}
\end{figure}
%

The effects of various physical factors on the density
dependence of the bremsstrahlung emissivity in the lattice
of ground-state matter at $T= 3 \times 10^8$ K is shown
in Fig.~\ref{figcomp}. We present the total
emissivity, and also the static-lattice and
phonon emissivities. In addition,
we show the total emissivity
calculated neglecting either the effects of band-structure,
the Debye--Waller factor, or the nuclear form factor.
Here by the phrase ``neglecting the Debye--Waller factor''
we mean the one-phonon approximation in which the Debye--Waller
factor is set to zero.
One can see that
the phonon contribution is noticeably smaller than the
static-lattice one, and the ratio of the static-lattice and
phonon contributions increases slowly with density, reaching
a factor of about 5 at $\rho = 10^{13}$ g cm$^{-3}$.
The effect of the form factor also increases with
density while the effect of the Debye--Waller factor
becomes lower.

\begin{figure}[ht]
\begin{center}
\vspace*{1ex}
\leavevmode
\epsfysize=8.5cm \epsfbox{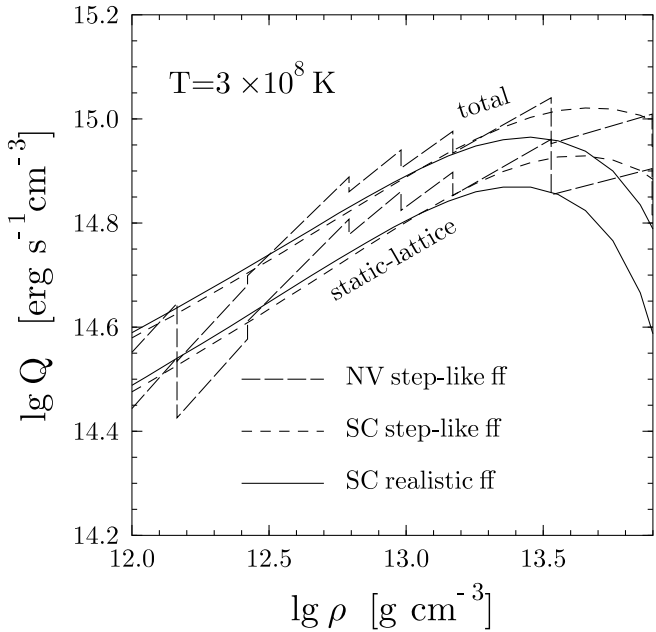}
\end{center}
\caption[ ]{
Density dependence of the bremsstrahlung emissivity
at $T=3 \times 10^8$ K for the ground-state matter
composed of spherical nuclei.
Long-dashed lines are calculated with the step-like
form factor (ff) given by Eq.~(\protect{\ref{form factor}}) 
for the nuclear composition by Negele \& Vautherin (NV,1973).
Short-dashed lines are calculated with the same form factor
but for the smooth-composition (SC) model of matter (Sect.\ 2).
Solid lines correspond to the SC model and more realistic form factor
for the proton distribution (\protect{\ref{Oya}}). Upper lines
show the total emissivities, while the lower ones give the static
lattice contribution alone
}
\label{fignvo}
\end{figure}

The neutrino emissivities presented in Figs.\ \ref{figfe}--\ref{figcomp}
are calculated with the simplified atomic form factor
(\ref{form factor}) appropriate for a step-like proton
distribution within the nuclei. We have verified that the
simplified form factor gives quite accurate results for
$\rho \la 10^{13}$ g cm$^{-3}$. However, at higher $\rho$
the neutrino emissivity becomes sensitive to the shape
of the proton distribution. Then the form factor
based on the proton-density distribution (\ref{Oya})
seems to be more reliable, as discussed in Sect.\ 2.
We have made a series of calculations with this
more realistic form factor making use of the smooth-composition
(SC) model of the ground-state matter (Sect.\ 2).
Figure \ref{fignvo} is an extension of Fig.~\ref{figcomp}
to higher densities ($10^{12}$ g cm$^{-3} \leq \rho \leq 8 \times 10^{13}$
g cm$^{-3}$) for matter containing spherical nuclei. The
highest density, $\rho = 8 \times 10^{13}$ g cm$^{-3}$,
is close to the transition to the phases with nonspherical nuclei.
We now compare the neutrino emissivities calculated with
the realistic and step-like form factors.
The realistic form factor is included in both
the phonon contribution (through Eq.\ (\ref{Lph}))
and the static-lattice contribution (Sect.\ 3.3).
The emissivities obtained with the realistic form factor
for the SC model are compared with those obtained with
the step-like form factor for the SC and Negele--Vautherin
models of matter. The total emissivities (upper curves)
are seen to be close to the static-lattice emissivities
(lower curves), which indicates that the static-lattice
contribution is dominant. The emissivities in the
Negele--Vautherin model show the familiar jumps
(e.g., Figs.\ \ref{figgro} and \ref{figcomp}) associated 
with variations of the nuclear composition.  After 
averaging over these jumps, they reproduce the emissivities 
derived in the SC model (with the same form factor). The 
emissivities obtained with the realistic and step-like form 
factors are seen to be close for $\rho \la 10^{13}$ g 
cm$^{-3}$.  However, when the density becomes higher than
$10^{13}$ g cm$^{-3}$ the realistic form factor decreases 
the electron-nucleus interaction and reduces noticeably the 
neutrino emission compared with the case of the step-like 
form factor.

\begin{figure}[ht]
\begin{center}
\vspace*{1ex}
\leavevmode
\epsfysize=8.5cm \epsfbox{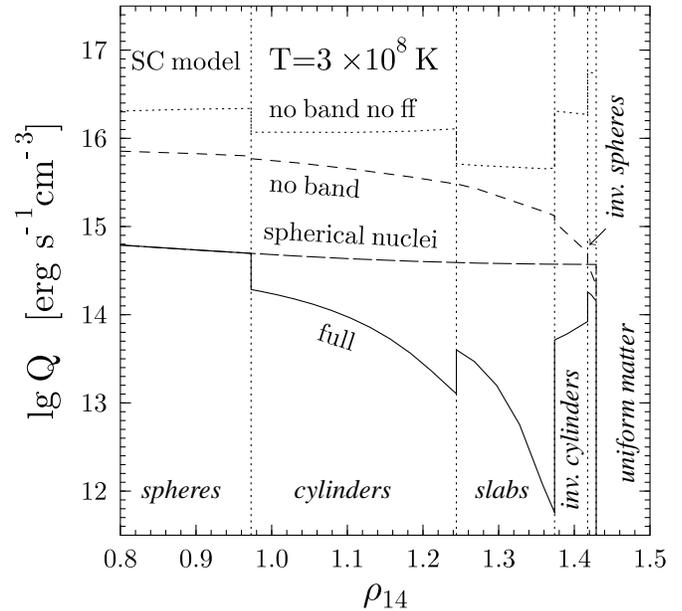}
\end{center}
\caption[ ]{
Density dependence of the bremsstrahlung  emissivity
at $T=3 \times 10^8$~K for five phases of spherical
and nonspherical atomic nuclei at the bottom of a neutron star crust
for the SC model of matter calculated with the realistic form factor.
Long-dash line is obtained
by assuming the nuclei to be spherical to the crust bottom.
Short-dash line is calculated for all phases neglecting
the band structure, while dotted line is obtained
neglecting the band structure and the form factor
}
\label{figedge}
\end{figure}

Figure \ref{figedge} is an extension of Figs.\ \ref{figgro} --
\ref{fignvo} towards higher densities.
It shows the density dependence
of the neutrino bremsstrahlung emissivities through
all five phases of spherical and nonspherical nuclei
for $T=3 \times 10^8$ K. In order to display all the phases we use a
linear density scale rather than a logarithmic
one; $\rho_{14}$ is the
density in units of $10^{14}$ g cm$^{-3}$. Densities in excess
of $1.43 \times 10^{14}$ g cm$^{-3}$ correspond to
uniform matter in the neutron star core. The emissivities
are calculated using the SC model.
In the spherical phase, the phonon contribution is
included and the Debye--Waller factor is taken into account.
In the nonspherical phases, we
neglect the Debye--Waller factor
and the phonon contribution (see above).
This circumstance is partly
responsible for the jumps in the neutrino emissivities
at $\rho = 9.73 \times 10^{13}$ g cm$^{-3}$, the
interface between the phases with spherical and cylindrical nuclei.
All the curves are calculated with the realistic form factor.
The long-dashed curve is obtained
assuming the nuclei remain spherical up to the highest densities at the
very
bottom of the neutron star crust and by including the phonon contribution
and the Debye--Waller factor. The parameters of
such nuclei are appropriate to the Negele--Vautherin
model of dense matter, smoothed over jumps.
We also present the emissivities
calculated for all phases of nonspherical nuclei
neglecting either the band-structure effects,
or both the band structure and the form factor.
The neutrino emission
at $\rho \sim 10^{14}$ g cm$^{-3}$
is very sensitive to the proton charge distribution.
A neglect of the form factor leads to
overestimation of the neutrino emissivity by 1 -- 1.5
orders of magnitude. The effects of nonspherical phases
are also rather important. Non-sphericity of the nuclei
mainly lowers the neutrino emission by reducing
the dimension of the sums over reciprocal lattice vectors
in Eq.\ (\ref{S}).
The reduction can exceed one order
of magnitude (cf the solid and dashed curves near the
cylinder--slab interface). More work
is required to calculate the Debye--Waller factor and
the phonon contribution and determine accurately the
bremsstrahlung emission for nonspherical nuclei.
However the spherical--nucleus model
probably represents a reasonable upper limit to this emissivity for
densities where the nuclei can be nonspherical.

\begin{figure}[ht]
\begin{center}
\leavevmode
\epsfysize=8.5cm \epsfbox{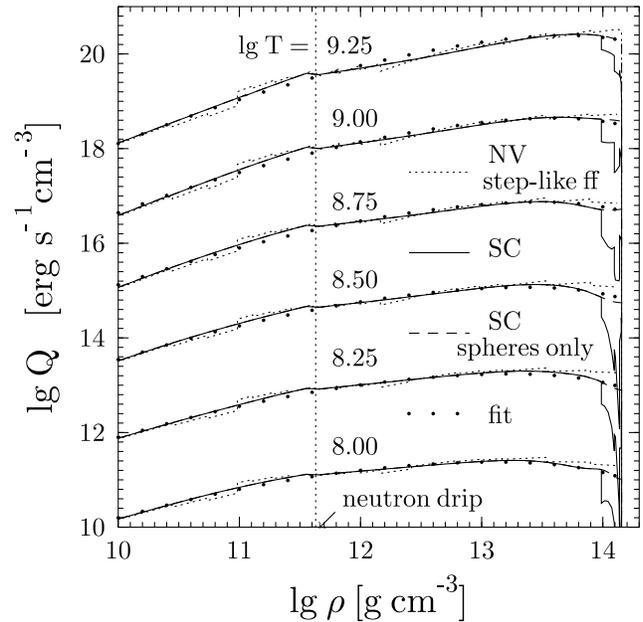}
\end{center}
\caption[ ]{
Density dependence of the bremsstrahlung  emissivity
for the ground-state matter of the neutron-star crust
at six temperatures $T$ in the model by Negele \&
Vautherin (NV, dots) (\cite{nv73})  with the step-like form factor (ff),
and in the SC model with the realistic form factor assuming either
the nuclei are spherical to the crust bottom (dashes)
or including non-spherical phases (solid lines).
Filled circles show our fits (\protect{\ref{Qfit}}) to dashed lines

}
\label{figtot}
\end{figure}
%

Figure \ref{figtot} shows the density dependence of neutrino
bremsstrahlung for the ground-state matter
of the neutron-star envelopes at six temperatures, from
$10^8$ K to $1.8 \times 10^9$ K, in the density range
$10^{10}$ g cm $^{-3}  \leq  \rho  \leq 1.4 \times 10^{14}$ g cm$^{-3}$.
The $\rho-T$ domain displayed is the most
important one for application to
neutron-star cooling.
The dotted curves are calculated using the
Negele--Vautherin model of matter and the step-like form factors
(\ref{form factor}).
The dashed lines are obtained for the SC model
with the realistic form factor and on the assumption that nuclei
are spherical to the bottom of the crust.
The solid lines are also derived for the SC model with the
realistic form factor but with allowance for
the phases of nonspherical nuclei.
Sharp decreases of the curves at $\rho \ga 10^{14}$
g cm$^{-3}$ occur because the nuclei become nonspherical
(see Fig.~\ref{figedge}).

Our calculations for spherical nuclei with the realistic form factor
in wide density and temperature ranges,
$10^9$ g cm$^{-3} \leq \rho \leq 1.5 \times 10^{14}$ g cm$^{-3}$
and $5 \times 10^7$ K $\leq T \leq 2 \times 10^9$ K
can be fitted by the expression
\begin{eqnarray}
&&  \!\!\!\!\!  
\lg Q \, [{\rm erg \, cm^{-3} \, s^{-1}}]  = 
         11.204+7.304 \, \tau+0.2976 \, r
\nonumber\\&&\qquad\qquad
     - 0.370 \, \tau^2  +  0.188 \, \tau r- 0.103 \,r^2 + 0.0547 \, \tau^2 r
   \nonumber\\&&\qquad\qquad
-6.77 \, \lg \left(1+ 0.228 \rho / \rho_0 \right),
\label{Qfit}
\end{eqnarray}
where $\tau = \lg T_8$, $r = \lg \rho_{12}$,
and $\rho_0 = 2.8 \times 10^{14}$ g cm$^{-3}$ is
the standard nuclear-matter density.
The relative error of this fit formula generally does not exceed 1\%
(in $\lg Q$) over the indicated $\rho-T$ domain.
The accuracy of the fit is seen from Fig.~\ref{figtot}.
The fitting formula reproduces the main features of the neutrino
bremsstrahlung radiation in a very wide
density domain $10^9$ g cm$^{-3} \leq \rho \leq 10^{14}$ g cm$^{-3}$,
where the atomic nuclei are expected to be spherical,
and it probably gives a realistic estimate of the emissivity
for higher densities, $10^{14}$ g cm$^{-3} \leq \rho \leq
1.4 \times 10^{14}$ g cm$^{-3}$, where the nuclei become nonspherical.
This fit formula should be quite sufficient for many applications.

\begin{figure}[ht]
\begin{center}
\leavevmode
\epsfysize=8.5cm \epsfbox{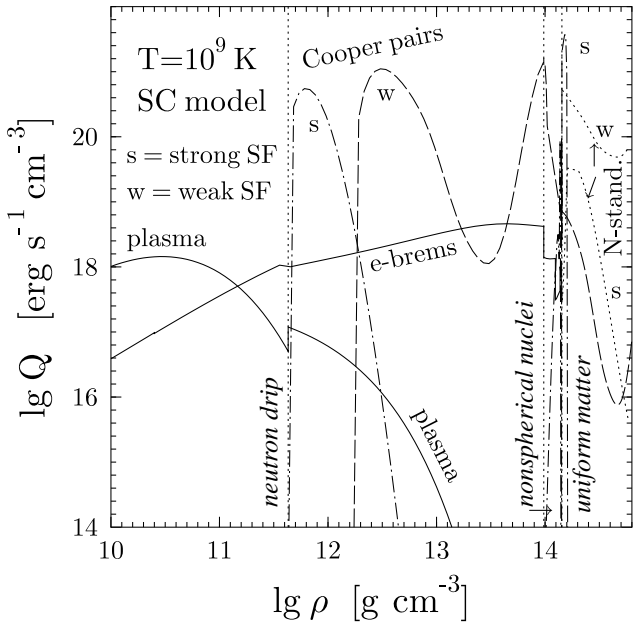}
\end{center}
\caption[ ]{
Density dependence of the neutrino emissivities produced
in a neutron star crust and core
at $T= 10^9$ K by various
neutrino generation mechanisms: electron bremsstrahlung (e-brems),
plasmon decay (plasma),
Cooper pairing of nucleons (Cooper pairs) in the models
of strong (s) and weak (w) neutron superfluidity (SF), and standard
neutrino generation mechanisms (N-stand.) in
strongly (s) and weakly (w) superfluid uniform matter (see text for
details)
}
\label{figt9}
\end{figure}

\begin{figure}[h]
\begin{center}
\leavevmode
\epsfysize=8.5cm \epsfbox{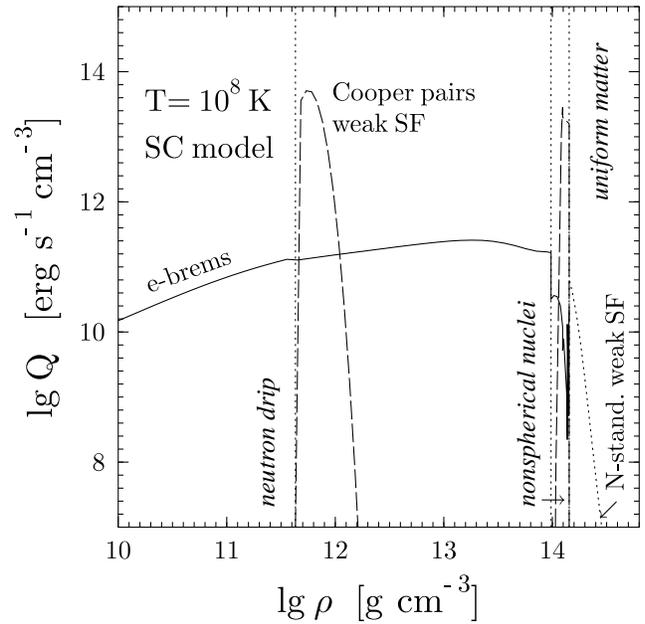}
\end{center}
\caption[ ]{
Same as in Fig.~\protect{\ref{figt9}} but for $T=10^8$ K.
Plasmon decay, standard neutrino
emission from uniform mater and Cooper-pairing emission
in the strongly superfluid matter
become negligible
\label{figt8}
}
\end{figure}

Finally, in Figs.\ \ref{figt9} and \ref{figt8}
we compare neutrino bremsstrahlung
by electrons in the neutron-star crust with other
neutrino emission mechanisms for $T=10^9$ K and $10^8$ K,
respectively. The other mechanisms considered are neutrino
emission due to plasmon decay (e.g., Itoh et al.\ \cite{itoh-ea96})
and due to the nucleon Cooper-pair formation under the action
of nucleon superfluidity (e.g., Yakovlev et al.\ \cite{ykl98}).
The SC model of ground-state matter is used.
In principle, some contribution at low $T$ may come
from neutrino emission due to scattering of electrons
by charged impurities (e.g., Haensel et al.\ \cite{hkya96}).
We ignore this mechanism here since it is determined
by impurity parameters which are largely unknown.
We show also the neutrino emission
produced by some mechanisms in
uniform matter of the neutron-star core:
the standard neutrino energy losses
and the nucleon Cooper-pair neutrinos
in superfluid uniform matter.
It is assumed that uniform matter has a moderately
stiff equation of state proposed by Prakash et al.\ (\cite{pal88})
(the same version as was used by Page \& Applegate \cite{pa92}).
The standard neutrino processes include the
modified Urca reactions and nucleon-nucleon bremsstrahlung.
All the standard reactions are partially
suppressed by the combined action of the neutron and
proton superfluids in uniform matter. The reaction rates
and suppression factors are taken in the form described by
Levenfish \& Yakovlev (\cite{ly96}). The proton and neutron
superfluid critical temperatures depend on
density. We assume singlet-state pairing of protons,
and either singlet-state or triplet-state pairing of
neutrons.
Neutron pairing is expected to occur in a singlet state
in matter at densities less than roughly that of
the core-crust interface,
and in a triplet state at higher densities.
For any density we adopt the type of neutron
pairing that corresponds to the higher critical temperature.
The strong density dependence of the neutrino emissivities
in uniform matter (Figs.\ \ref{figt9} and \ref{figt8})
is due to the pronounced density dependence of
the neutron and proton critical temperatures at $\rho \sim \rho_0$.
Since critical temperatures are very sensitive
to the microscopic model of the nucleon interaction, we have
considered two cases, corresponding to strong (s) and weak (w)
nucleon superfluids (SFs). The strong superfluid model
is based on the rather large superfluid gaps
calculated by Elgar{\o}y et al.\ (\cite{eehjo96b}) for singlet-state pairing
(with maximum gap of about 2.5 MeV as a function
of nucleon number density) and by Hoffberg et al.\ (\cite{hgrr70})
for triplet-state pairing. The weak superfluid model
makes use of the small superfluid gaps
derived by Wambach et al.\ (\cite{wap93})
(with a maximum gap of about 1 MeV) for the singlet
superfluid and by Amundsen \& {\O}stgaard (\cite{ao85b})
for the triplet neutron superfluid.  For singlet pairing we regard the
weak pairing case to be the more realistic because the calculations of
Wambach et al.\ (\cite{wap93}) included the effects of induced 
interactions.

The mechanism of neutrino production due
to Cooper pairing of nucleons was proposed
by Flowers et al.\ (\cite{frs76})
and independently by Voskresensky \& Senatorov (\cite{vs86}, \cite{vs87})
for uniform matter. A critical comparison of these
works has been done by Yakovlev et al.\ (\cite{ykl98})
who considered also the case of triplet neutron
pairing.
The theory predicts a powerful maximum of
the Cooper-pair neutrino emission when the temperature falls
below the critical temperature $T_{\rm c}$ for neutrons
or protons. However, at still lower temperatures,
$T\ll T_{\rm c}$, the emission falls exponentially.
The emission is much stronger for
neutrons than for protons due to the smallness
of axial-vector electroweak currents for protons
(Yakovlev et al.\ \cite{ykl98}).
In spite of its long history, this process was
forgotten for a long time, and it has been included
in cooling simulations only recently
(Page \cite{p97}, \cite{p98}; Schaab et al.\ \cite{schaab-ea97}; 
Levenfish et al.\ \cite{lsya98};
Yakovlev et al.\ \cite{ykl98}).
We make use of the results by
Yakovlev et al.\ (\cite{ykl98}) for uniform matter and
for the crust. In the uniform matter, we take into account
neutrino emission due to pairing of neutrons and protons.
In the crust, the Cooper neutrino emission is evaluated
including the contribution from free nucleons alone
(from free neutrons in all phases
of matter in the inner crust, and from free protons in the phases
with cylinders and spheres of neutrons surrounded by nuclear matter).
In principle, there can be a substantial
contribution (Yakovlev et al.\ \cite{ykl98}) from the non-uniform distribution
of the nucleons within atomic nuclei (Sect.\ 2).
We neglect this effect here but we intend to discuss
it in a separate article.

In a high temperature plasma, $T = 10^9$ K,
at $\rho \la 10^{11}$ g cm$^{-3}$ (Fig.~\ref{figt9}),
the process most competitive with electron bremsstrahlung
is the well-known plasmon decay
(Itoh et al.\ \cite{itoh-ea96}). However its rate
falls exponentially with decreasing $T$,
and the process almost dies out at $T=10^8$ K (Fig.~\ref{figt8}).
The standard neutrino emission from uniform matter is
greatly reduced by the nucleon superfluidity. It decreases
exponentially when the temperature becomes much smaller
than the superfluid critical temperatures of nucleons. For instance,
the standard emission is quite substantial at $T=10^9$ K
but becomes much less significant at $T=10^8$ K
(practically negligible for the case of strong superfluid).
Neutrino emission by Cooper pairing in the uniform matter
and in the inner crust
is also exponentially suppressed when the temperature is
much lower than the critical temperatures of
neutrons and protons. Accordingly the temperature and
density dependence of the Cooper-pair neutrino emissivity is very strong.
If $T=10^9$ K and the superfluidity is strong,
we have two peaks of Cooper-pair neutrinos: one near the
neutron drip point (at $\rho \sim 10^{12}$ g cm$^{-3}$),
and a very narrow peak near the core-crust interface
($\rho \approx 1.4 \times 10^{14}$ g cm$^{-3}$).
Both peaks are produced by neutron pairing, and
are pronounced since the neutron critical temperature
is sufficiently small (only slightly exceeds $T$)
in the indicated density ranges
even for strong superfluidity. The smallness of $T_{\rm c}$
in the first density range is associated with a
low number density of free neutrons near the neutron drip point,
and the smallness in the second density range corresponds to
the transition from singlet to triplet neutron pairing.
Outside these density ranges the neutron critical temperature
is too high and the emissivity of the Cooper neutrinos
is exponentially suppressed. When the temperature
decreases the emissivity becomes smaller, and the
process dies out at $T=10^8$ K (cf.\ Figs.\ \ref{figt9} and \ref{figt8}).

If $T=10^9$ K and the superfluidity is weak, the Cooper
pairing appears to be the dominant neutrino emission
process in a large fraction of the neutron star crust
since the neutron critical temperature is not much higher
than $T$. However, the process is practically switched
off at low densities $\rho \la 2 \times 10^{12}$ K, because
a weak neutron superfluid has not yet occurred
at these densities ($T_{\rm c} < T=10^9$ K). When the temperature drops
to $10^8$ K, the neutrino emission due to Cooper
pairing is suppressed. Nevertheless, two high
peaks of the emissivity survive (similar to those
for a strong superfluid at $T=10^9$ K). The first one
corresponds to low $\rho \sim 10^{12}$ g cm$^{-3}$
($T_{\rm c}$ is not much higher than $10^8$ K), and the second
corresponds to $\rho \sim 1.4 \times 10^{14}$ g cm$^{-3}$, where
$T_{\rm c}$ is low, which corresponds 
to the transition from a singlet to a triplet neutron superfluid.

We conclude that the main contribution to
neutrino emission from a neutron-star crust
comes from two processes, the neutrino-pair bremsstrahlung
and Cooper pairing of neutrons. The bremsstrahlung
neutrino emission has been calculated rather reliably,
excluding possibly in the phases of nonspherical nuclei
near the core-crust interface.
The mechanism operates in a wide ranges of densities and
temperatures, and the density dependence of the emissivity
is generally smooth. The neutrino emission due to the
Cooper pairing of nucleons is extremely sensitive
to the model adopted for the superfluid gaps in the nucleon spectra.
This mechanism is more important for lower gaps
(weaker superfluid). The emissivity can be a sharp
function of density and temperature.
We remark that the microscopic models that correspond to weak
superfluidity are likely the more reliable since they incorporate the
effects of induced processes (screening) in the effective
neutron-neutron interaction.  In addition, in the phases with non-spherical
nuclei, the neutron superfluid gap in the matter is expected to be reduced
by the presence of nuclei, in which the matter has a higher density, and
a smaller pairing interaction,  than in the neutrons outside nuclei. We
expect to consider the Cooper-pair process in more detail in a
future publication.

%
\section{Conclusions}
We have analysed the neutrino pair emission (\ref{brems}) 
due to the bremsstrahlung of degenerate relativistic 
electrons at densities from $10^{9}$ g cm$^{-3}$ to $1.5 
\times 10^{14}$ g cm$^{-3}$ and temperatures from $5 \times 
10^7$ to $5 \times 10^9$ K in the neutron-star crusts.  We 
have presented the expressions for the neutrino emissivity 
from a plasma of liquid and solid atomic nuclei taking into 
account the effects of finite sizes (the nuclear form 
factor) of the nuclei.  In solid matter, we have studied 
the static-lattice and phonon contributions 
to the neutrino bremsstrahlung 
with allowance for the electron band-structure 
and multi-phonon processes.  
We have considered bcc, fcc and hcp Coulomb
crystals, and showed that the neutrino emission is insensitive to the
lattice type.  We have made use of two models of matter in the neutron star
crusts:  ground-state matter and accreted matter.  We have proposed a
smooth-composition model of ground-state matter to analyse the neutrino
bremsstrahlung near the bottom of the crust, where the shapes of the local
nucleon density distributions over Wigner--Seitz cells become important.
This smooth-composition model can be applied also for calculating
the electron transport properties (thermal and electric conductivities)
in the deep layers of the crust.
We have calculated the static-lattice contribution for nonspherical nuclei
at the very crust bottom, at densities $\rho \ga 10^{14}$ g cm$^{-3}$.
We
have analyzed (Sect.\ 4) the neutrino emissivity as a function of density,
temperature, and nuclear composition.  We have obtained a simple analytic
fit (\ref{Qfit}) to the bremsstrahlung emissivity for ground-state matter
composed of spherical nuclei in the neutron-star crusts.  We expect the
fit to give reliable values of the neutrino bremsstrahlung emissivity for
densities $\rho \leq 10^{14}$ g cm$^{-3}$, where the nuclei are certainly
spherical, and to give a reliable upper limit of the 
emissivity for $10^{14}$ g cm$^{-3} \leq \rho \leq 1.4 
\times 10^{14}$ g cm$^{-3}$, where the nuclei become 
nonspherical.

Our results can be used, for instance, to study cooling of isolated neutron
stars.  A not too old star (of age lower than about $10^5$ -- $10^6$ yr)
cools mainly via neutrino emission from its interior (e.g., Pethick \cite{p92}).
The main contribution to the neutrino luminosity comes usually from the
stellar core.  However, neutrino emission from the crust can also be
important.  In young neutron stars it plays a significant role
in thermal relaxation of the stellar interiors.  It can also be
important in rather old neutron stars during the transition from the
neutrino
cooling stage to the photon one.  The neutrino luminosity of the
stellar core decays generally somewhat faster than the luminosity of the
crust, and the crustal luminosity survives longer.  Moreover, the neutrino
emission from the crust can dominate the emission from the core in the
neutron stars with highly superfluid cores and/or in stars with a stiff
equation of state.  In the latter case, the crust can be quite massive and
its neutrino luminosity can be substantial.  Finally, low-mass neutron
stars always possess relatively massive crusts, and their neutrino
luminosities can be mainly determined by their crusts.

A number of problems remain to be solved in connection with the calculation
of neutrino pair bremsstrahlung in the crust.  One of these is the
influence on the phonon spectrum of neutrons outside nuclei when matter is
made up of spherical nuclei.  Intuitively one would expect the neutron
liquid to make the effective mass of nuclei larger, thereby decreasing
phonon frequencies.  This in turn would increase the bremsstrahlung from
phonon processes at least at low temperatures.
Another is the nature of the collective
excitations of the phases with non-spherical nuclei.  There are a number of
these that have low frequencies, and these will give a``phonon
contribution", while at the same time affecting the static lattice
contribution.

\begin{acknowledgements}
The authors are grateful to D.\ Baiko for many helpful
discussions. 
This work was supported by RFBR (grant 96-02-16870a),
RFBR-DFG (grant 96-02-00177G),
INTAS (grant 96-0542), the US National Science Foundation
(grant NSF AST96-18524), and NASA (grant NAGW-1583).
\end{acknowledgements}


\end{document}